\documentclass[reprint,onecolumn,superscriptaddress,preprintnumbers,nofootinbib,prd,amsmath,amssymb]{revtex4-2}
\pdfoutput=1 
\usepackage{graphicx}
\usepackage{dcolumn}
\usepackage{bm}
\usepackage{adjustbox}
\usepackage{multirow}
\usepackage{slashed}
\usepackage{hhline}
\usepackage{color}
\usepackage[usenames, dvipsnames]{xcolor}
\usepackage[utf8]{inputenc}
\usepackage{tabularx,booktabs,caption}
\usepackage{hyperref}
\hypersetup{
    pdfnewwindow=true,      
    colorlinks=true,       
    linkcolor=blue,          
    citecolor=blue,        
    filecolor=blue,      
    urlcolor=blue           
}

\usepackage{ulem}
\usepackage{marginnote}
\usepackage{cancel}
\usepackage{amssymb}
\usepackage{textcomp}
\usepackage{mathrsfs}
\usepackage{feynmp-auto}
\usepackage{subcaption}
\usepackage{ulem}
\usepackage{empheq}
\usepackage{framed}
\usepackage{float}
\usepackage[utf8]{inputenc}

\usepackage{bbm,dsfont}
\usepackage{setspace}
\usepackage{mathrsfs}
\usepackage[small]{titlesec}
\usepackage[multidot]{grffile}
\usepackage[position=b]{subcaption}
	\makeatletter
	\let\@@magyar@captionfix\relax
	\makeatother

\newcommand{\pfrac}[2]{\left(\dfrac{#1}{#2}\right)}

\newcommand{\ket}[1]{\left\lvert #1\right\rangle}
\newcommand{\bra}[1]{\left\langle #1\right\rvert}
\newcommand{\be}{\begin{equation}}
\newcommand{\ee}{\end{equation}}
\newcommand{\bea}{\begin{eqnarray}}
\newcommand{\eea}{\end{eqnarray}}
\newcommand{\lsim}{\buildrel < \over {_\sim}}

\newcommand{\eq}{\mathrm{(eq)}}
\newcommand{\BL}{{B-L}}

\allowdisplaybreaks

\newcommand{\onbb}{{0\nu\beta\beta}}
\newcommand{\TeV}{\mathrm{TeV}}
\newcommand{\GeV}{\mathrm{GeV}}

\newcommand{\TUM}{\affiliation{Physik Department T70, James-Franck-Stra{\ss}e, Technische Universit\"at M\"unchen, 85748 Garching, Germany}}
\newcommand{\PRISMA}{\affiliation{PRISMA+ Cluster of Excellence \& Mainz Institute for Theoretical Physics, FB 08 - Physics, Mathematics and Computer Science, Johannes Gutenberg-Universit\"at Mainz, 55099 Mainz, Germany}}
\newcommand{\UMASS}{\affiliation{Amherst Center for Fundamental Interactions, Department of Physics, University of Massachusetts, Amherst, MA 01003 USA}}
\newcommand{\TDLI}{\affiliation{Tsung-Dao Lee Institute and School of Physics and Astronomy, Shanghai Jiao Tong University, 800 Dongchuan Road, Shanghai, 200240 China}}

\newcommand{\CALTECH}{\affiliation{Kellogg Radiation Laboratory, California Institute of Technology, Pasadena, CA 91125 USA}}
\newcommand{\INT}{\affiliation{Institute for Nuclear Theory, University of Washington, Seattle, WA 98195-1550, USA}}
\newcommand{\NU}{\affiliation{Northwestern University, Department of Physics \& Astronomy, 2145 Sheridan Road, Evanston, IL 60208, USA}}

\begin{document}

\preprint{ACFI-T21-08, TUM-HEP-1346/21}

\title{TeV-scale Lepton Number Violation: Connecting Leptogenesis, Neutrinoless Double Beta Decay, and Colliders}

\author{Julia Harz}
\email{julia.harz@uni-mainz.de}
\TUM\PRISMA

\author{Michael J. Ramsey-Musolf}
\email{mjrm@sjtu.edu.cn, mjrm@physics.umass.edu}
\TDLI\UMASS\CALTECH

\author{Tianyang Shen}
\email{tysimonshen@gmail.com}
\UMASS

\author{Sebasti\'an Urrutia Quiroga}
\email{suq90@uw.edu}
\UMASS\INT\NU

\begin{abstract}
In the context of TeV-scale lepton number violating (LNV) interactions, we illustrate the interplay between leptogenesis, neutrinoless double beta ($0\nu\beta\beta$) decay, and LNV searches at proton-proton colliders. Using a concrete model for illustration, we overcome the limitations of previous EFT analyses and are able to identify the parameter space where standard thermal leptogenesis is rendered unviable due to washout processes. Moreover, we show how $0\nu\beta\beta$ decay and $pp$ collisions provide complementary probes. We find that the new particle spectrum can have a decisive impact on the relative sensitivity of these two probes.
\end{abstract}

\maketitle

\section{Introduction}

In the Standard Model (SM) of particle physics, the lepton number (L) is accidentally conserved at the classical level yet not conserved at the quantum level due to the B$+$L anomaly. The corresponding energy scale, associated with the electroweak sphaleron, is $E_\mathrm{WS}\sim 10$ TeV. When embedding the SM in a more complete theory, it is straightforward to introduce explicit lepton number violating (LNV) interactions. The associated BSM LNV mass scale $\Lambda$ may {\it a priori} range anywhere from well below $E_\mathrm{WS}$ to the grand unification scale, $M_\mathrm{GUT}$.\\

Explaining the origin of neutrino mass provides guidance for the choice of scale, though it need not be definitive. The well-known type I seesaw mechanism \cite{Minkowski:1977sc, Gell-Mann:1979vob, Yanagida:1979as, Glashow:1979nm, Mohapatra:1979ia}, for example, suggests $\Lambda\lsim M_\mathrm{GUT}$. In this minimal extension of the SM with right-handed neutrinos (RHN),  
a lepton-number-conserving Dirac mass term would lead to massive neutrinos, as neutrino oscillations require. However, unless one explicitly requires lepton number conservation, one may also include an LNV RHN mass term. The corresponding Majorana mass $m_N$ sets the LNV scale $\Lambda$. Diagonalization of the full mass matrix for the neutral leptons implies that that light neutrinos are also Majorana particles. For $m_N$ within a few orders of magnitude of $M_\mathrm{GUT}$, the Dirac mass Yukawa couplings may be as large as $\mathcal{O}(1)$ while accommodating the scale of light neutrino masses implied by neutrino oscillations and cosmological neutrino mass bounds. The light neutrino interactions then inherit the LNV properties of Majorana neutrinos.\\

It is entirely possible that the dynamics responsible for neutrino mass generation entail a scale well below the conventional seesaw scale. Indeed, a variety of well-motivated scenarios involving LNV at the TeV scale and below have been widely considered. More generally, one may encounter BSM LNV interactions at multiple scales, some of which may give the dominant contribution to light neutrino masses $m_\nu$ with others playing a less significant role. In the R-parity violating (RPV) minimal supersymmetric Standard Model (MSSM), for example, $m_\nu$ can arise at tree- or loop-level through, respectively, bilinear or trilinear RPV terms in the superpotential. Introducing RHN superfields allows one to generate $m_\nu$ through the conventional, high-scale see-saw mechanism while including much lower-scale LNV RPV contributions.\\

In what follows, we consider the phenomenological consequences of this last possibility, namely, that LNV interactions are realized in nature at multiple scales, including the conventional seesaw scale and scales commensurate with or below $E_\mathrm{WS}$. The array of possible consequences is rich, including observations from cosmology, searches for new phenomena in high-energy collisions, and tests of fundamental symmetries in low-energy, high-sensitivity experiments. A number of earlier studies have considered the implications of multiple-scale LNV at one or more of these frontiers, such as the search for neutrinoless double $\beta$ decay ($0\nu\beta\beta$). In fact, the possibility of TeV-scale LNV contributions to the latter process, in addition to the \lq\lq standard mechanism\rq\rq{} involving the virtual exchange of three light Majorana neutrinos, has become a topic of considerable interest for the $0\nu\beta\beta$ community (see Refs. \cite{Cirigliano:2022oqy, Acharya:2023swl} and references therein). Moreover, there exists a considerable body of work focusing on probing the TeV scale and below LNV with proton-proton ($pp$) and $e^+e^-$ collisions, as well as several studies considering the interplay of these probes with $0\nu\beta\beta$. Recently, the authors of Ref. \cite{Li:2020flq} analyzed the interplay of $0\nu\beta\beta$ and cosmological bounds on the sum of light neutrino masses in this context. In a similar vein, the authors of Refs. \cite{Deppisch:2015yqa, Deppisch:2017ecm, Li:2019fhz, Deppisch:2020oyx} pointed out the significant implications of multiple-scale LNV for the viability of baryogenesis via leptogenesis using an effective field theory (EFT) framework.\\

To our knowledge, there exists little, if any, work that provides an integrated analysis of all three frontiers under the multiple-scale LNV paradigm. Here, we endeavor to do so using a simplified model framework. We seek to address several questions:
\begin{itemize} 
\item If multiple-scale BSM LNV is realized in nature, under what conditions (model setup and parameter choices) could it be discovered experimentally?
\item If such a scenario is identified, under what conditions would it either allow for or preclude leptogenesis? 
\item What aspects of the LNV interactions are most important for the viability of leptogenesis, and how sensitive are terrestrial experiments to these aspects?
\end{itemize}

Addressing these questions requires carrying out detailed studies of the early universe leptogenesis dynamics, the sensitivity of the LHC and future colliders to BSM LNV interactions, and state-of-the-art computations of the $0\nu\beta\beta$ rate. Because of the multiple energy scales involved in these processes, including temperature $T$ in the early universe, the high energy collider center of mass energy $E_\mathrm{com}$, and the nuclear and hadronic scales $M_\mathrm{nuc}$ and $M_\mathrm{had}$ pertinent to nuclear transitions, performing an integrated study for all three frontiers requires adopting a framework valid in all three energy regimes. Choosing a specific, ultraviolet (UV) complete model would certainly satisfy this requirement, though with the loss of generality engendered by model-specific details and phenomenological constraints. On the other hand, when the multiple-scale LNV setup includes the TeV scale, reliance on an EFT would not be appropriate for consideration of LHC and future collider studies. Thus, we will employ a simplified model that embodies some features of several UV complete scenarios, fully aware that it might not be realized in nature in an of itself. We hope, nevertheless, to provide - with some degree of generality -- a template for future, integrated inter-frontier studies using either simplified or complete models that seek to address the above-mentioned questions.\\

Simplified models have been used extensively in other contexts, such as dark matter studies \cite{SimplMod:DM1, SimplMod:DM2, SimplMod:DM3}, electroweak phase transition dynamics \cite{SimplMod:EWPT1, SimplMod:EWPT2, SimplMod:EWPT3}, and collider phenomenology \cite{SimplMod:Coll1, SimplMod:Coll2, SimplMod:Coll3}, to mention a few examples. Our particular simplified model choice is intended to highlight the connection between $\onbb$ decay and collider phenomenology in light of the viability of thermal leptogenesis in a spirit of broadness and generality. In the context of analyzing $\onbb$ decay, this model choice is inspired by the earlier work in Refs. \cite{Peng:2015haa, Li:2021fvw} where it gives rise to the leading-order (LO) long-range pion-exchange amplitude expected to have the maximal impact on the $\onbb$-decay rate.\footnote{Among the subset of simplified models that share this feature, the one that we adopt here minimally extends the SM in terms of particles and interactions. See Refs. \cite{Prezeau:2003xn, Graesser:2016bpz}, and references therein.} 
In this work we include the new interactions explicitly in the leptogenesis Boltzmann equations with thermal mass effects taken into account, allowing us to analyze in detail the dependence of the BAU on the masses and couplings associated with the new particles and their interactions. From the Boltzmann equation solutions, we identify the regions of the model mass and coupling parameter space for which the TeV scale LNV interactions would render unviable standard thermal leptogenesis (also assuming the presence of the heavy RHN as described above). We then utilize state-of-the-art hadronic and nuclear physics methods relevant to  $0\nu\beta\beta$ decay and machine learning techniques for collider LNV searches to delineate the sensitivity of these probes to the leptogenesis unviable parameter space. In the context of our illustrative model, we find that
\begin{itemize}
\item The observation of an LNV signal at the Large Hadron Collider (LHC) and/or a future 100 TeV $pp$ collider would preclude the viability of standard thermal leptogenesis.
\item The observation of $0\nu\beta\beta$ decay could also rule out the standard leptogenesis paradigm, assuming the $0\nu\beta\beta$ decay amplitude is dominated by the TeV scale LNV mechanism. Additional information, such as the results from $pp$ collider LNV searches, knowledge of the light neutrino mass hierarchy, and/or the sum of light neutrino masses would be needed to identify the underlying $0\nu\beta\beta$ decay mechanism.
\item The relative reaches of $0\nu\beta\beta$ decay and collider LNV searches depend decisively on the new particle spectrum — a feature not readily seen within the previously used pure EFT approach.
\item The observation of experimental signature would be consistent with the scale of light neutrino masses implied by neutrino oscillation experiments as well as cosmological and astrophysical neutrino mass probes.
\end{itemize}

The outcome of our analysis leading to these conclusions is organized as follows: In Section~\ref{sec:model}, we first introduce the simplified model set-up that we used for our study. Then, we introduce the Boltzmann-equation framework and the consequences with respect to leptogenesis in Section~\ref{sec:leptogenesis}. In Section~\ref{sec:collider}, we discuss the performed collider analysis, and in Section~\ref{sec:nldbd}, the treatment with respect to $0\nu\beta\beta$ decay. In Section~\ref{sec:discussion}, we present our results and conclude in Section~\ref{sec:conclusions}.
\section{A simplified model for TeV-scale LNV}
\label{sec:model}

For the standard leptogenesis scenario \cite{Fukugita:1986hr}, at least two heavy right-handed neutrinos $N_j$ are needed to generate a CP-asymmetry via the one-loop decay of the lightest right-handed neutrino \cite{Nanopoulos:1979gx, Kolb:1979qa}. We consider the broadly studied situation \cite{Davidson:2008bu, Buchmuller:2004nz, Giudice:2003jh} where the lepton asymmetry is produced in a single flavor, the neutrino masses are hierarchical ($m_{N_1}\ll m_{N_2},m_{N_3}$), and the decays of the two heavier neutrinos ($N_2$ and $N_3$) are neglected. Henceforth, we will refer to the lightest right-handed neutrino $N_1$ simply as $N$, dropping the flavor subscript. The interaction part of the Lagrangian is given by:
\begin{equation}
\mathscr{L} = y_N\,\overline{L}(i\tau^2)H^\ast N-\frac{m_N}{2}\overline{N^c}N + \mathrm{h.c.}\ ,
\label{eq:LagN}
\end{equation}

where $L=(\nu_L, e_L)^\top$ and $\tau$'s are the Pauli matrices in isospace. Notice that in the context of standard thermal leptogenesis, consistency with light neutrino phenomenology implies $m_N > 10^9$ GeV \cite{Davidson:2002qv}. While assuming that a lepton-asymmetry might have been generated via the decay of right-handed neutrinos at a high scale, we want to investigate the impact of additional LNV interactions at the TeV scale. For these purposes, we adopt a simplified model framework that has been previously used to explore the $\onbb$-decay and collider interplay \cite{Peng:2015haa, Li:2021fvw}. This particular model represents a possible realization of the dim-9 effective operator studied in Refs.~\cite{Prezeau:2003xn, Deppisch:2015yqa, Graesser:2016bpz} as we will discuss in more detail later.\\

As discussed in Refs. \cite{Peng:2015haa, Li:2021fvw}, one possible minimal model that gives rise to the LO $\pi\pi ee$ interactions responsible for $\onbb$ decay\footnote{As discussed in Refs. \cite{Prezeau:2003xn, Graesser:2016bpz, Cirigliano:2017ymo, Cirigliano:2018yza}, several quark-lepton effective operators can give rise to different interactions in the chiral Lagrangian. See Ref. \cite{Graesser:2022nkv} for an example of a simplified model mapping onto $\pi NNee$ interactions.} (see Section \ref{sec:nldbd} for a detailed discussion) includes a scalar $S$ transforming as $(1,2,1)$ under $SU(3)_C \times SU(2)_L \times U(1)_Y$ and a Majorana fermion $F$ that transforms as an SM gauge singlet\footnote{Note, we use the convention $Y=Q_\mathit{EM}-T_3$.}. The Lagrangian reads
\begin{equation}
\widetilde{\mathscr{L}} = g_Q\,\overline{Q}Sd_R +  g_L\,\overline{L}(i\tau^2)S^\ast F-m_S^2\,S^\dagger S - \frac{m_F}{2}\,\overline{F^c}F + \lambda_{HS}\,(S^\dagger H)^2 + \mathrm{h.c.} + ... \ ,
\label{eq:ModelO2}
\end{equation}

where $Q=(u_L,d_L)^\top$ and $q_R=(u_R,d_R)^\top$ are the left-handed and right-handed quark isospinors, respectively. In a full, UV-complete theory such as RPV SUSY \cite{Allanach:2003eb}, $S$ and $F$ are identified as the slepton and the lightest neutralino fields, respectively. The ellipsis in Eq. \eqref{eq:ModelO2} indicates other possible terms such as $SH^3$ and $S^3H$. For simplicity, we will omit those terms and also assume that the heavy neutrino $N$ will not interact with the new fields introduced. It is important to notice that our simplified model assumes $\langle S\rangle=0$ at the tree level. Although the scalar potential is, in principle, arbitrary and a positive $S$-mass term could accommodate a zero VEV, these assumptions are made for the sake of simplicity since their effects are not necessarily related to our focus on the $\onbb$-decay/collider interface.\\

\begin{figure}
\centering
 \includegraphics[scale=0.8]{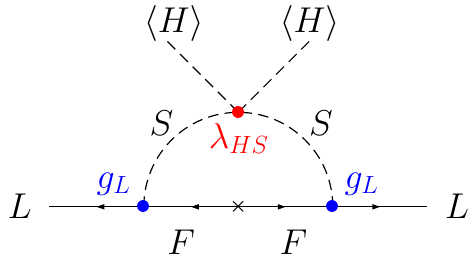} \\
 \caption{One-loop contribution to the Weinberg operator \cite{Weinberg:1979sa} induced by the interactions in Eq. \ref{eq:ModelO2}. Note that the magnitude of this contribution is proportional to the coupling $\lambda_{HS}$ that does not enter the amplitudes for $0\nu\beta\beta$-decay or same-sign dilepton plus di-jet production in $pp$ collisions.}
 \label{fig:O2_mass}
\end{figure}

Besides the generation of small Majorana neutrino masses via the see-saw mechanism \cite{Minkowski:1977sc, Yanagida:1980xy, Glashow:1979nm, GellMann:1980vs, Mohapatra:1980yp} induced by the right-handed neutrino $N$, 

\begin{equation}
m_\nu^{\rm type-I}\sim \frac{y_N}{m_N}v^2\,,
\label{eq:vmass_seesaw}
\end{equation}

where $v$ is the vacuum expectation value (VEV) of $H$, additional contributions can be generated at the one-loop level via the interactions in the Lagrangian~\eqref{eq:ModelO2}, as shown in Fig. \ref{fig:O2_mass}. By denoting $m_{R,I}$ to the mass of the real and imaginary parts of the neutral component $S^0$ of the doublet $S$, the neutrino mass contribution from our simplified model is given by\footnote{Due to the similarity, in terms of particle content, between our simplified model and the Scotogenic model \cite{DeRomeri:2021yjo}, it is possible to use those calculations for comparison.}
\begin{equation}
m_\nu^{\rm simpl\ mod} \sim \frac{g_L^2}{2(4 \pi)^{2}}\,m_F\left[\frac{m_{R}^{2}}{m_{R}^{2}-m_{F}^{2}} \log \left(\frac{m_{R}^{2}}{m_{F}^{2}}\right)-\frac{m_{I}^{2}}{m_{I}^{2}-m_{F}^{2}} \log \left(\frac{m_{I}^{2}}{m_{F}^{2}}\right)\right]\,,
\label{eq:vmass}
\end{equation}
where $m_{R,I}^2=m_S^2\pm \lambda_{HS}\,v^2$. In the limit of $\lambda_{HS}\ll1$,
\begin{equation}
m_\nu^{\rm simpl\ mod} \sim \frac{\lambda_{HS}}{m_F}\frac{g_L^2\,v^2}{(4 \pi)^{2}}\left[\frac{m_F^2}{m_S^2-m_F^2} - \frac{m_F^4}{\big(m_S^2-m_F^2\big)^2} \log \left(\frac{m_S^2}{m_F^2}\right)\right]\,,
\end{equation}
which matches the naïve loop-counting estimation in Ref. \cite{Cai:2017jrq}. If $S$ acquires a non-zero VEV, the mixing between $S$ and $H$ will modify the estimation in Eq. \eqref{eq:vmass}. Additionally, the total neutrino mass $m_{\nu}$ is given by the combinations of the expressions in Eqs. \eqref{eq:vmass_seesaw} and \eqref{eq:vmass}, where $m_\nu=m_\nu^{\rm type-I}+m_\nu^{\rm simpl\ mod}$. In certain parameter regions, one of these two contributions may be more dominant than the other. We will not address either of these possibilities here since we focus on illustrating the interplay between colliders and $\onbb$-decay experiments. In practical terms, we will study the model in the limit of $\lambda_{HS}\to0$.\\

For low energy $0\nu\beta\beta$-decay process, the heavy particles in the Lagrangian~\eqref{eq:ModelO2} can be integrated out, yielding the effective dim-9 LNV interaction:
\begin{equation}
\mathscr{L}_\mathit{LNV}^\mathrm{eff} = \frac{C_1}{\Lambda^5}\mathcal{O}_1 +\mathrm{h.c.} \  , \quad  \mathcal{O}_1 =\bar{Q}\tau^+d\bar{Q}\tau^+d\bar{L}L^c \,.
\label{eq:eq3}
\end{equation}

We can match
\begin{equation}
C_1=g_L^2\,g_Q^2 \quad\mathrm{and}\quad \Lambda^5 = m_S^4\,m_F\ .
\label{eq:matching}
\end{equation}

Interestingly, this demonstrates that TeV-scale masses for $m_S$ and $m_F$ are not in conflict with constraints from neutrino masses,~\eqref{eq:vmass}, as in such a model realisation the contribution to $\onbb$-decay (depending on $g_L, g_Q$ only) is independent from lowest order contribution to the neutrino mass (depending on $\lambda_{HS}$). In the following, we will study the impact of the new interactions in~\eqref{eq:ModelO2} on the baryon asymmetry generated from the heavy-right handed neutrinos and their detection possibilities at colliders and $\onbb$-decay experiments.
\begin{figure}
\centering
	\includegraphics[scale=0.8]{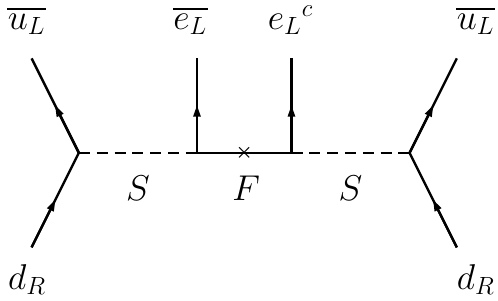} \\
 \caption{The realization of the $0\nu\beta\beta$-decay dim-9 operator induced by the interactions in Eq. \ref{eq:ModelO2}. The scalar $S$ transforms as $(1,2,1)$ under $SU(3)_C \times SU(2)_L \times U(1)_Y$, and the Majorana fermion $F$  transforms as an SM gauge singlet. }
 \label{fig:onbb_diagram}
\end{figure}


\section{Leptogenesis}
\label{sec:leptogenesis}
LNV interactions can play an important role in generating the baryon asymmetry of the Universe (BAU), which is usually quantified in terms of the baryon-to-photon number density based on the PLANCK 2018 data~\cite{Zyla:2020zbs, Aghanim:2018eyx} 
\begin{align}
\eta_B^{\mathrm{obs}} = \frac{n_B}{n_\gamma} = (6.12\pm 0.04)\times 10^{-10}\,,
\label{eq:BAUobs_eta}
\end{align}
or the yield, normalized to the entropy density $s$,
\begin{align}
Y_B^{\mathrm{obs}} = \frac{n_B}{s} = (8.71\pm 0.06)\times 10^{-11}\,.
\label{eq:BAUobs_Y}
\end{align}
As a key test of standard cosmology, this value agrees with limits coming from Big Bang Nucleosynthesis~\cite{Zyla:2020zbs}. According to the established three Sakharov conditions \cite{Sakharov:1967dj}, a mechanism explaining the BAU needs to include (i) baryon number (B-) violation, (ii) C- and  CP-violation and (iii) an out-of-equilibrium condition (or CPT-violation). As these conditions are not sufficiently satisfied within the SM, BSM physics is required.\\

One of the most popular explanations for the observed baryon asymmetry is baryogenesis via leptogenesis \cite{Fukugita:1986hr}. In this mechanism, a 
lepton asymmetry is generated via the $B-L$ violating decays of right-handed heavy neutrinos. Due to the interference of the tree-level and one-loop contribution to the decays and the presence of at least two right-handed neutrinos, a net $CP$-asymmetry can occur. When the decay falls out of equilibrium during the cooling of the Universe, a final lepton asymmetry is generated. If this happens before the electroweak phase transition, the Standard Model electroweak sphaleron processes can transfer this lepton asymmetry into a baryon asymmetry. In the standard leptogenesis scenario, the same interactions that induce the right-handed neutrino decay also cause $\Delta L = 1$ and $\Delta L = 2$ scattering processes that can destroy again this asymmetry, the so-called ``washout'' processes. If the latter are too strong, the generated asymmetry can be again destroyed. In \cite{Deppisch:2013jxa}, it was shown that the observation of a generic $\Delta L = 2$ lepton number violating signal at the LHC or in $\onbb$-decay experiments (via an operator of dimension seven or higher) would directly imply a significant washout rate and hence would render the asymmetry generation insufficient~\cite{Deppisch:2015yqa, Deppisch:2017ecm}. While this interplay has been previously described in an effective field theory approach only, we want to investigate this within our simplified model set-up as described in Section~\ref{sec:model}. To this end, we analyze the potential to generate the observed baryon asymmetry within our simplified model, which consists of the SM extended by a right-handed neutrino (standard leptogenesis scenario) and an additional new physics contribution as defined in Eq.~\eqref{eq:ModelO2} leading to additional LNV washout processes experimentally accessible at the TeV scale. \\

\begin{figure}[t]
	\centering
	\begin{subfigure}[b]{0.33\textwidth}
		\centering
            \includegraphics[width=0.6\textwidth]{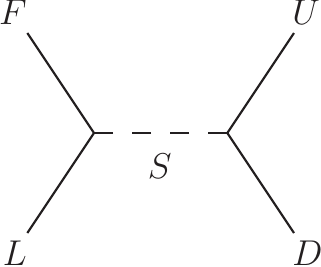}
		\caption{$\widetilde{S}_s$ with $\Delta L=1$}
	\end{subfigure}%
	~ 
	\begin{subfigure}[b]{0.33\textwidth}
		\centering
		\includegraphics[width=0.35\textwidth]{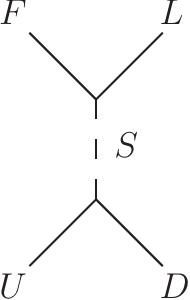}
		\caption{$\widetilde{S}_t$ with $\Delta L=1$}
	\end{subfigure}%
	\begin{subfigure}[b]{0.33\textwidth}
		\centering
		\includegraphics[width=0.35\textwidth]{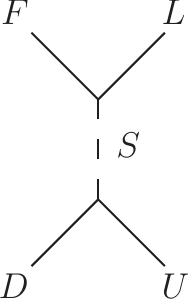}
		\caption{$\widetilde{S}_t$ with $\Delta L=1$}
	\end{subfigure}
	
		\vspace{1.2cm}
	\begin{subfigure}[b]{0.33\textwidth}
		\centering
            \includegraphics[width=0.6\textwidth]{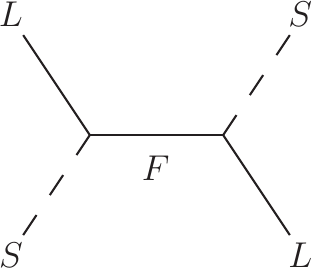}
		\caption{$\widetilde{N}_s$ with $\Delta L=2$}
	\end{subfigure}%
	~ 
	\begin{subfigure}[b]{0.33\textwidth}
		\centering
		\includegraphics[width=0.35\textwidth]{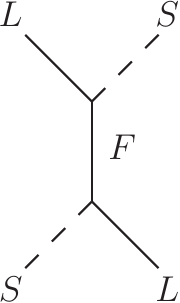}
		\caption{$\widetilde{N}_s$ with $\Delta L=2$}
	\end{subfigure}%
	~
	\begin{subfigure}[b]{0.33\textwidth}
		\centering
		\includegraphics[width=0.35\textwidth]{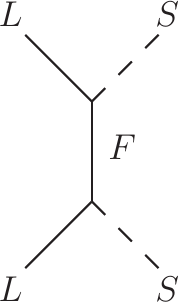}
		\caption{$\widetilde{N}_t$ with $\Delta L=2$}
	\end{subfigure}
\caption{Relevant lepton-number violating processes arising from our extended Lagrangian that contribute to the washout additionally to the usual washout processes which occur within the standard leptogenesis scenario including right-handed neutrinos.}
\label{fig:LNVdiagrams}
\end{figure}

First, we classify and study the relevant processes that will have an impact on the predicted baryon asymmetry. Hereby, we indicate the contributions that arise from Eq.~\eqref{eq:ModelO2}, with a tilde (~$\widetilde{}$~) and the contributions arising from the standard thermal leptogenesis Lagrangian, Eq.~\eqref{eq:LagN}, without any additional marker (see Fig.~\ref{fig:LNVdiagrams}):

\begin{itemize}
	\item Decays and inverse decays ($\Delta L=1 $):
	\begin{align}
	D &\equiv [N \leftrightarrow H^\pm L^\mp] \,. \label{eq:ovdecayN} \\
	\widetilde{D} &\equiv [F \leftrightarrow S^\pm L^\mp] \,.  
	\end{align}
	
	\item Scattering processes ($\Delta L=1 $), with the subscript indicating exchange via $s$- or $t$-channel:
	\begin{align}
	S_s &= [LN \leftrightarrow U_3D_3]\quad,\qquad 2S_t = [N\bar{U}_3 \leftrightarrow D_3\bar{L}] + [N\bar{D}_3 \leftrightarrow U_3\bar{L}] \,.\\
	\widetilde{S}_s &= [LF \leftrightarrow U_1D_1]\quad,\qquad 2\widetilde{S}_t = [F\bar{U}_1 \leftrightarrow D_1\bar{L}] + [F\bar{D}_1 \leftrightarrow U_1\bar{L}] \,.
	\end{align}
	
	\item Scattering processes ($\Delta L=2 $) with the Majorana fermion $F$ as mediator:
	\begin{align}
	N_s &\equiv [LH \leftrightarrow \bar{L}\bar{H}] \qquad\textrm{and}\qquad N_t \equiv [LL \leftrightarrow \bar{H}\bar{H}]\,. \\
	\widetilde{N}_s &\equiv [LS \leftrightarrow \bar{L}\bar{S}] \qquad\textrm{and}\qquad \widetilde{N}_t \equiv [LL \leftrightarrow \bar{S}\bar{S}]\,. \label{eq:ovscatt2O2}
	\end{align}
\end{itemize}

Hereby, the subscript $s$($t$) depicts a corresponding $s$($t$)-channel exchange as indicated in Fig.~\ref{fig:LNVdiagrams}. In our analysis, we generally neglect scattering processes with gauge bosons in order to avoid an unphysical logarithmic enhancement of $g^2 \ln(gT/m_N)$ that would not occur in a full thermal calculation at NLO. In principle, IR divergences would arise for a soft gauge boson. If regularized with a thermal mass, a logarithmic dependence ($g^2 \ln(gT/m_N)$) would remain. As was demonstrated in \cite{Salvio:2011sf} for the standard leptogenesis scenario with right-handed neutrinos, this logarithmic dependence would cancel when including the corresponding thermal distribution of the gauge bosons in a thermal plasma. This cancellation would, for instance, naturally occur in a calculation using the Closed Time Path formalism \cite{Garbrecht:2018mrp}. As the scattering processes can be generally seen as a higher-order correction to the corresponding decays, it was recommended by \cite{Salvio:2011sf} not to take these problematic processes into account, as a regularization with a thermal mass would lead to bigger uncertainties than when directly neglecting those. As a similar behavior is expected for the gauge scattering in our extended model (the dominant washout contribution arises from the inverse decay, as will be discussed later in more detail), we also neglect these processes in our washout calculation. Following the same argumentation, we also neglect the scattering with light quarks that would give rise to IR divergences in the standard leptogenesis scenario. As IR divergences do not appear in the quark scattering in our extended model set-up due to the massive scalar $S$ in the propagator (cp. Fig.~\ref{fig:LNVdiagrams} (a)-(c)), we systematically include these scattering processes in our washout calculation. In follow-up work, it would be desirable to include these kinds of processes consistently via a non-equilibrium approach in quantum field theory, such as the closed-time path formalism \cite{Garbrecht:2018mrp} or an effective approach as suggested in \cite{Blazek:2021gmw, Blazek:2022azr}. Given that these processes are expected to lead to more washout, our results can be interpreted as conservative with respect to the washout strength.\\

To describe the evolution of the lepton asymmetry, we include all the outlined processes, Eqs. \eqref{eq:ovdecayN}-\eqref{eq:ovscatt2O2} and derive the Boltzmann equations for the yield of the right-handed neutrino $Y_N=n_N/s$ and the ($B-L$) asymmetry $Y_{\BL}$. Notice that the yield of the ($B-L$) asymmetry $Y_{B-L}$ is related with the yield of the baryon asymmetry $Y_{B}$ by \cite{Chen:2007fv}
\begin{align}
 Y_{B}=\frac{28}{79}Y_{B-L}\,
\end{align}

For example, for the evolution of  $Y_N$, we can write
\begin{equation}
zHs \frac{dY_N}{dz} = -\sum_{a,i,j} [Na \leftrightarrow ij]\,,
\label{original_BE}
\end{equation}
with $z=m_{N0}/T$ being the dimensionless time parameter, $m_{N0}$ is the zero-temperature mass, and $s$ the entropy density. Hereby, $[Na \leftrightarrow ij]$ comprises all relevant processes and permutations:
\begin{eqnarray}
[Na \leftrightarrow ij] &=& \left[ \dfrac{n_Nn_a}{n_N^\eq n_a^\eq } - \dfrac{n_in_j}{n_i^\eq n_j^\eq }\right] \gamma^\eq(ij \leftrightarrow Na)\,,
\end{eqnarray}
with $\gamma^\eq(ij \leftrightarrow Na)$ indicating the corresponding equilibrium reaction rates. In the derivation, we neglected Pauli blocking or Bose enhancement, which is to a good accuracy valid ($\langle E \rangle \approx 3T$ such that $(1 \pm f) \approx 1$ \cite{Strumia:2006qk}). The expression for the $2-2$ scattering processes is given by
\begin{equation}
\gamma^\eq_S(ij\leftrightarrow ab) = \dfrac{T}{64\pi^4}\int_{s_{\min}}^\infty ds\ \sqrt{s}\,\hat\sigma(s)\,K_1\left(\dfrac{\sqrt{s}}{T}\right)\,,
\end{equation}
with the reduced cross section $\hat\sigma \equiv 2s\,\lambda[1,m_i^2/s,m_j^2/s]\,\sigma$. Hereby, $\sigma$ is the total cross section summed over initial and final states, $\lambda[a,b,c]$ the K\"allen function, and $s$ the square of the center-of-mass energy. The corresponding relation for decays reduces to
\begin{equation}
\gamma^\eq_D(N \leftrightarrow ij\cdots) = n_N^\eq\dfrac{K_1(z)}{K_2(z)}\Gamma_N \qquad\textrm{with}\quad n_N^\eq = \dfrac{3\zeta(3)g_NT^3}{8\pi^2}z^2K_2(z)\,,
\label{eq:gammaDeq}
\end{equation}
with $n_N^\eq$ being the equilibrium number density, $\Gamma_N$ the decay width, and $g_N=2$ the number of degrees of freedom. Collecting all relevant processes that change the abundance of the neutrino, $Y_N$, and the one of the ($B-L$)-asymmetry, $Y_\BL$, we finally arrive at two coupled Boltzmann equations 
\begin{eqnarray}
\dfrac{dY_N}{dz} &=& -(D + S)\Big(Y_N - Y_N^\eq\Big) \label{BE1}\\
\dfrac{dY_\BL}{dz} &=& -\epsilon\,D\,\Big(Y_N - Y_N^\eq\Big) - W^{\mathrm{tot}}\,Y_\BL \label{BE2}\,.
\end{eqnarray}
We assume that all SM particles are in thermal equilibrium at all times. Due to its fast gauge interactions, we can assume equilibrium also for the new, heavy particle $S$. The new particle $F$, however, is not strictly in equilibrium during all of the relevant time. We have checked that the equilibrium assumption will not affect the evolution of $Y_{\BL}$ up to $z=10^6$, while having a significant advantage concerning the computing time. While it may be reasonable to extrapolate this result up to $z=10^8$, we leave explicit verification to future work when more substantial computing resources become available. Furthermore, $D,~W$, and $S$ indicate the relevant (inverse) decay, washout, and scattering processes, including both the usual standard leptogenesis interactions as well as the ones arising from our new Lagrangian. We define the usual washout parameter $K$ as
\begin{align}
 K = \frac{\Gamma_N}{H(z=1)}\,.
\end{align}
In the following, we will study both the strong ($K=10^{2}$) and weak ($K=10^{-2}$) washout scenario.
As the new contribution in our model, Eq.~\eqref{eq:ModelO2} is not directly interacting with the sterile neutrino sector, the Boltzmann equation for the change of the heavy neutrino number density remains unaltered with respect to the standard leptogenesis case. The decay rate $D$ is given by
\begin{align}
 D(z) = \dfrac{\gamma_{D}}{zH(z)n_N^\eq} = \dfrac{1}{zH(z)} \dfrac{\,K_1(z)}{K_2(z)} \Gamma_D\,,
 \label{eq:decayrateN}
\end{align}
with $H(z)=1.66 (\sqrt{g_*}m^2_{N0})/(M_{\mathrm{Pl}}\,z^2)$ and $K_{n}(z)$ being the modified Bessel function of the second kind.
The scattering rates, $S = 2S_s + 4S_t$, in Eq.~\eqref{BE1} are given as
\begin{align}
\label{eq:SN}
S_{s/t}(z) = \dfrac{\gamma_{S_{s/t}}}{zH(z)n_N^\eq} = \dfrac{1}{zH(z)} \dfrac{m_N}{48 \pi^2 \zeta(3) K_2(z)} I_{S_{s/t}}(z) \,,
\end{align}
with 
\begin{align}
\label{eq:ISN}
I_{\{S_{s/t}, N\}}(z) = \int_{x_{\min}}^\infty dx\ \sqrt{x}\,\hat\sigma_{\{S_{s/t}, N\}}(x)\,K_1(\sqrt{x}\,z)\,.
\end{align}
For the simplification of some expressions, we have introduced the parameter $x=s / m_N^2$. In these expressions, $m_N$ differs from $m_{N0}$ because it will have a $T$-dependent component when thermal effects are included.\\

\begin{figure}[t]
	\centering
	\includegraphics[width=0.6\textwidth]{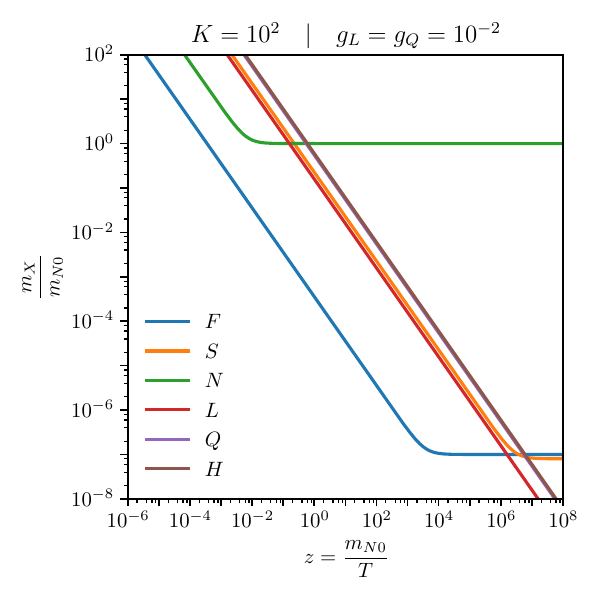}\\
	\caption{Evolution of the thermal masses. Due to a negligible difference between the thermal masses of $Q$, $u$, and $d$, we only show $m_Q$. In this regime, the thermal masses are basically independent of the initial $m_X(T=0)$ value. This is in contrast to the thermal mass of the right-handed neutrino whose thermal correction leaves the large $m_{N0}=10^{10}~\mathrm{GeV}$ unchanged until small values of $z<10^{-2}$. }
	\label{fig:thermalmasses}
\end{figure}

In contrast, the Boltzmann equation describing the $Y_{B-L}$-asymmetry evolution has to be adjusted to incorporate the new processes involved. However, as we do not assume any other $CP$ violating source than the heavy neutrino decay, the decay rate $D$ in the $Y_\BL$ Boltzmann evolution remains unaltered. However, for the washout, we have to consider both the standard washout contributions and our new interactions such that we adjust the washout term $W$ as follows
\begin{align}
W^{\mathrm{tot}}(z) = W(z, Y_N) + \widetilde{W}(z)\,.
\end{align}
The contribution $W(z, Y_N)$ corresponds to the expression of the standard case 
\begin{align}
 W(z, Y_N) = 2\Big[N(z)+S_t(z)\Big]\pfrac{Y_N^\eq}{Y_\BL^\eq} + S_s(z)\pfrac{Y_N}{Y_\BL^\eq} \,,
 \label{eq:Wstd}
\end{align}
with the $\Delta L = 1$ scattering rates $S_{s/t}$ as defined in Eqs.~\eqref{eq:SN} \eqref{eq:ISN}, and the $\Delta L = 2$ scattering $N(z)=N_{s}(z)+N_{t}(z)$ given as
\begin{align}
N(z) = \dfrac{\gamma_{N}}{zH(z)n_N^{\mathrm{(eq)}}} = \dfrac{1}{zH(z)} \dfrac{m_N}{48 \pi^2 \zeta(3) K_2(z)} I_{N}(z)\,,
\label{eq:NN}
\end{align}
with the integral $I_{N}(z)$ stated in Eq.~\eqref{eq:ISN}. Note that the inclusion of real intermediate states (RIS) in the $N_s$ scattering can lead to double counting with respect to the decays of the heavy right-handed neutrinos. Hence, as suggested in \cite{Salvio:2011sf}, we do not include the decay term in the washout contribution in order to prevent double counting.
The new contributions to the washout can be expressed as
\begin{align}
 	\widetilde{W}(z) = \left[\frac{1}{2}\widetilde{D}(z) + 2 \widetilde{N}(z)+ 2 \widetilde{S}_t(z)+ \widetilde{S}_s(z) \right]\pfrac{Y_N^\eq}{Y_\BL^\eq}\,,
 	\label{eq:WO2}
\end{align}
where the decay rate of the heavy particles $F$ or $S$, depending on their mass hierarchy, is given by
\begin{align}
 \widetilde{D}(z) = \dfrac{\gamma_{\widetilde{D}} }{zH(z)n_N^{\mathrm{(eq)}}}= \dfrac{1}{zH(z)} f_{F,S} \left(\dfrac{m_{F,S}}{m_N} \right)^2 \dfrac{\,K_1(z\,m_{F,S} / m_N)}{K_2(z)} \Gamma_{F,S}\,.
 \label{eq:decayrateO2}
\end{align}
Comparing with the decay rate of the heavy neutrino in Eq.~\eqref{eq:decayrateN} and the interaction rate in Eq.~\eqref{eq:gammaDeq}, we have to perform small adjustments. First, we rescaled the argument of the Bessel function by $(m_{F,S}/m_N)$. Secondly, we accounted for the equilibrium number density of $F$ or $S$ in $\gamma_{\widetilde{D}}$. Due to the following relation of the equilibrium number densities
\begin{eqnarray}
	n_F^\eq(z) &=& n_N^\eq(z)\,\pfrac{m_F}{m_N}^2\dfrac{K_2(z\,m_F/m_N)}{K_2(z)} \nonumber \,,\\
	n_S^\eq(z) &=& n_N^\eq(z)\,\dfrac{2}{3} \pfrac{m_S}{m_N}^2 \dfrac{K_2(z\,m_S/m_N)}{K_2(z)}\nonumber \,,
\end{eqnarray}
we have to rescale our expression by $f_{F,S}(m_{F,S}/m_N)^2$ with $f_F=1$ or $f_S=2/3$ due to the different number of degrees of freedom of $F,S$ with respect to $N$. Similarly, we can proceed with the scattering rates in Eqs.~\eqref{eq:SN} and \eqref{eq:NN} to adjust for the new physics contributions,
\begin{align}
\widetilde{S}_{s/t}(z) = \dfrac{\gamma_{\widetilde{S}_{s/t}}}{zH(z)n_N^{\mathrm{(eq)}}} = \dfrac{1}{zH(z)} \dfrac{m_N}{48 \pi^2 \zeta(3) K_2(z)} \left(\dfrac{m_F}{m_N} \right)^3 I_{\widetilde{S}_{s/t}}(z)\,,
\end{align}
\begin{align}
\widetilde{N}(z) = \dfrac{\gamma_{\widetilde{N}}}{zH(z)n_N^{\mathrm{(eq)}}} = \dfrac{1}{zH(z)} \dfrac{m_N}{48 \pi^2 \zeta(3) K_2(z)} \left(\dfrac{m_F}{m_N} \right)^3 I_{\widetilde{N}}(z)\,,
\end{align}
with
\begin{align}
I_{\{\widetilde{S}_{s/t}, \widetilde{N}\}}(z) = \int_{y_{\min}}^\infty dy\ \sqrt{y}\,\hat\sigma_{\{\widetilde{S}_{s/t}, \widetilde{N}\}}(y)\,K_1(\sqrt{y}\,z\, m_F/m_N)\,.
\end{align}
For convenience, we have defined $y = \sqrt{s}/m_F^2$. Note that, in contrast to the standard leptogenesis scenario, we do not include the RIS subtraction when the hierarchy $m_S > m_F$ holds (see Fig.~\ref{fig:thermalmasses}). Hence, no double counting occurs for $z\lsim10^6$ as the contribution $\widetilde{D}$ in Eq.~\ref{eq:WO2} includes only the inverse decays $FL \rightarrow S$ and not $SL \rightarrow F$. Only the latter one would lead to a double counting as it is already accounted for in the scattering processes $\widetilde{N}_{s}$ when the $F$ in the propagator is produced resonantly (cp. Fig.~\ref{fig:LNVdiagrams}~(d)). For later times, RIS subtraction is once again performed, and Eq.~\ref{eq:WO2} loses its inverse decay term (cf. Eq.~\ref{eq:Wstd}).\\

\begin{figure}[t]
\centering
\begin{subfigure}{0.5\textwidth}
    \centering
    \includegraphics[width=0.9\textwidth]{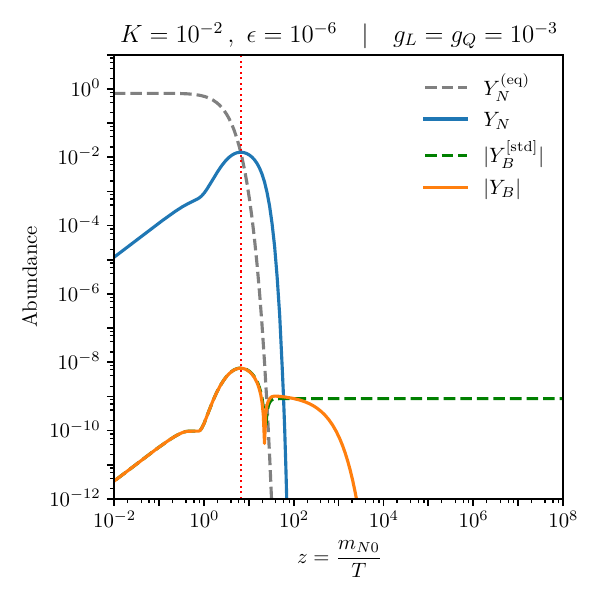}\\
    \caption{}
    \label{fig:evolutionDensity_a}
\end{subfigure}%
\begin{subfigure}{0.5\textwidth}
    \centering
    \includegraphics[width=0.9\textwidth]{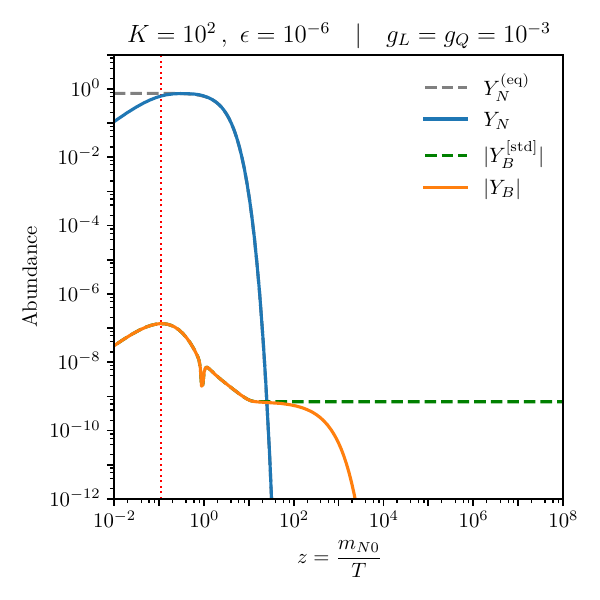}\\
    \caption{}
    \label{fig:evolutionDensity_b}
\end{subfigure}
\par\bigskip
\begin{subfigure}{0.5\textwidth}
    \centering
    \includegraphics[width=0.9\textwidth]{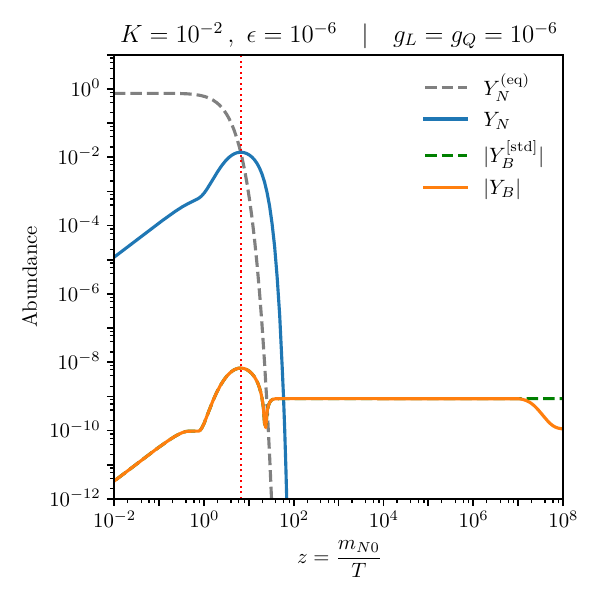}\\
    \caption{}
    \label{fig:evolutionDensity_c}
\end{subfigure}%
\begin{subfigure}{0.5\textwidth}
    \centering
    \includegraphics[width=0.9\textwidth]{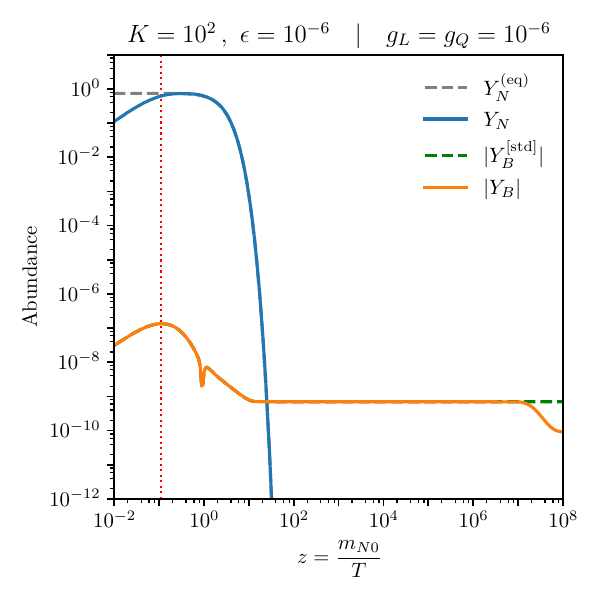}\\
    \caption{}
    \label{fig:evolutionDensity_d}
\end{subfigure}
\caption{Evolution of the yield of the right-handed neutrino $Y_N$ (blue solid line) and the yield of the baryon asymmetry for the standard scenario $Y^{[\mathrm{std}]}_B$ (green dashed line) and including our new contributions $Y_B$ (orange solid line) for the weak (left panels) and strong (right panels) washout regime and two different example values for $g_L=g_Q= \{10^{-3}, 10^{-6}\}$. The equilibrium abundance $Y_N^{\mathrm{(eq)}}$ is given as a gray dashed line. The red dotted line indicates $z_{\mathrm{eq}}$.}
\label{fig:evolutionDensity}
\end{figure}

In our set-up, we consider a heavy neutrino with $m_{N0} = 10^{10}~\mathrm{GeV}$ around which scale the generation of the asymmetry takes place roughly. As at such high temperatures, masses can receive sizeable thermal corrections, which could even lead to an altered mass hierarchy with respect to $T=0$. Hence, we consider thermal masses in our evolution of the neutrino and lepton number density:
\begin{align}
m_N^2(T) &= m^2_{N0} +\frac{1}{8} y^2_N T^2\\
m_F^2(T) &= m^2_{F0} + \frac{1}{8} g^2_L T^2 \label{eq:mF(T)}\\
m_S^2(T) &= m^2_{S0} +\left[\frac{3}{16}g_2^2 + \frac{1}{16}g_Y^2 + \frac{3}{12}g_Q^2 + \frac{1}{12}g_L^2\right] T^2 \label{eq:mS(T)}\\
m_H^2(T) &= \left[\frac{3}{16}g_2^2 + \frac{1}{16}g_Y^2 + \frac{1}{4}y_t^2 + \frac{1}{2}\lambda_H \right] T^2 \label{eq:mh(T)}\\
m_Q^2(T) &= \left[\frac{1}{6}g_3^2 + \frac{3}{32}g_2^2 + \frac{1}{288}g_Y^2 + \frac{1}{16}y_t^2 + \frac{1}{16}g_Q^2 \right] T^2\\
m_u^2(T) &= \left[\frac{1}{6}g_3^2 + \frac{1}{18}g_Y^2 + \frac{1}{8}y_t^2 \right] T^2\\
m_d^2(T) &= \left[\frac{1}{6}g_3^2 + \frac{1}{72}g_Y^2 + \frac{1}{8}y_b^2 +   \frac{1}{8}g_Q^2 \right] T^2\\
m_L^2(T) &= \left[\frac{3}{32}g_2^2 + \frac{1}{32}g_Y^2 +   \frac{1}{16}g_L^2 \right] T^2\,,
\end{align}
where we can express $T$ as $T=m_{N0}/z$. The definitions of $m_{F0}$ and $m_{S0}$ are analogous to $m_{N0}$. Here, we neglect contributions from the right-handed neutrino to the standard model masses, as those will only become relevant for $T > m_N$. In this regime, however, contributions are negligible with respect to the ones originating from the SM itself due to a comparably small coupling ($y^{\mathrm{strong}}_N=0.02$, $y^{\mathrm{weak}}_N=0.0002$). Hence they can be safely neglected.
We show the evolution of the thermal masses in Fig.~\ref{fig:thermalmasses}, where we have chosen $K=10^{2}$ such that $y_N=0.02$ and $g_L=g_Q=10^{-2}$. Relative to the heavy neutrino mass at zero temperature, the thermal corrections have almost no impact on $m_N(T)$, except for $z>10^{-2}$. This is in contrast to the evolution of the other particle masses. For instance, even when choosing at zero temperature $m_{F0}>m_{S0}$, the thermal corrections grow faster for $S$ than $F$ such that in the relevant temperature regime, the hierarchy of the particles changes (e.g., at $z=1$, $m_S>m_F$). Another interesting feature happens for the mass hierarchy of the Higgs boson and the right-handed neutrino. For $z \gtrsim 0.6$, the Higgs boson becomes heavier than the right-handed neutrino such that at higher temperatures, the decay $H \rightarrow NL$ opens up. We account for this effect by adapting $\Gamma_D$ in Eq.~\eqref{eq:decayrateN} to be $\Gamma_D=\Gamma (N \rightarrow LH)$ for $m_N(T)>m_H(T)$ and $\Gamma_D=\Gamma(H \rightarrow LN)$ for $m_N(T)<m_H(T)$. Moreover, the evolution of both the right-handed neutrino $Y_N$ and the baryon asymmetry $Y_B$ yields depend on the inclusion of thermal masses; our results in Fig.~\ref{fig:evolutionDensity} differ from previous zero-temperature calculations (cf. Ref.~\cite{Buchmuller:2004nz}) not only in thermal history but also in the final value of the asymmetry.\\

\begin{figure}[t]
\centering
\begin{subfigure}{0.5\textwidth}
    \centering
    \includegraphics[width=0.9\textwidth]{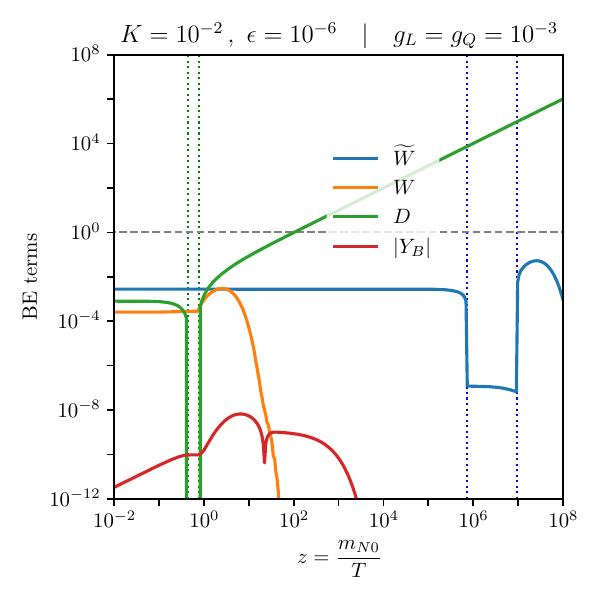}\\
    \caption{}
    \label{fig:evolutionContribs_a}
\end{subfigure}%
\begin{subfigure}{0.5\textwidth}
    \centering
    \includegraphics[width=0.9\textwidth]{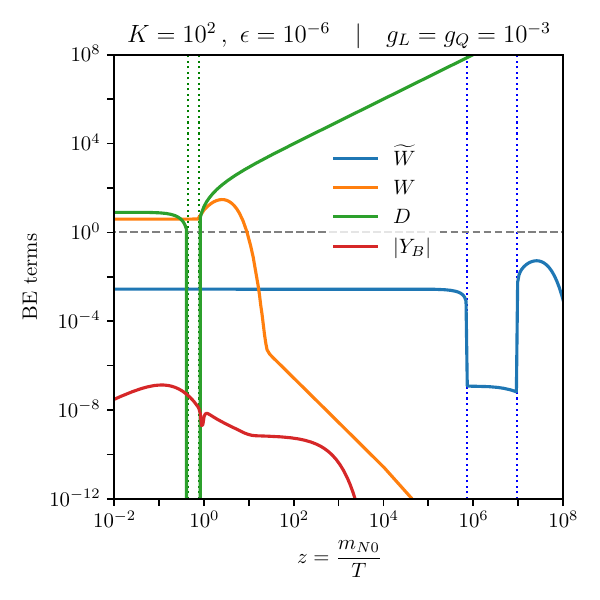}\\
    \caption{}
    \label{fig:evolutionContribs_b}
\end{subfigure}
\par\bigskip
\begin{subfigure}{0.5\textwidth}
    \centering
    \includegraphics[width=0.9\textwidth]{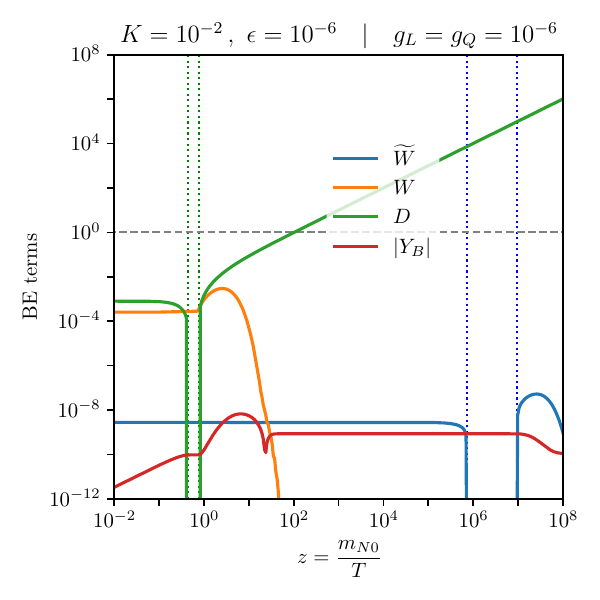}\\
    \caption{}
    \label{fig:evolutionContribs_c}
\end{subfigure}%
\begin{subfigure}{0.5\textwidth}
    \centering
    \includegraphics[width=0.9\textwidth]{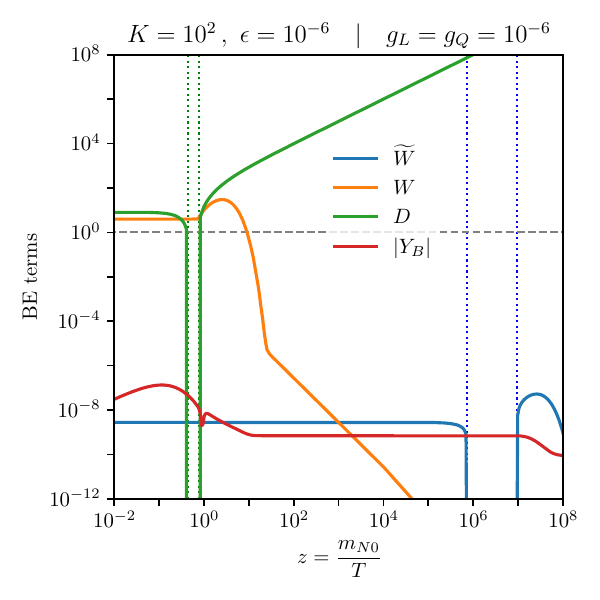}\\
    \caption{}
    \label{fig:evolutionContribs_d}
\end{subfigure}
\caption{Evolution of the different contributions relevant for the yield of the baryon asymmetry $Y_B$ (orange solid line) for the weak (left panels) and strong (right panels) washout regime and two different example values for $g_L=g_Q= \{10^{-3}, 10^{-6}\}$. The green dotted lines show when $m_H(z) = m_N(z)+m_L(z)$ and $m_N(z) = m_H(z)+m_L(z)$, respectively. The blue dotted line shows when $m_S(z) = m_F(z)+m_L(z)$.}
\label{fig:evolutionContribs}
\end{figure}

In order to study the impact of the additional contributions of our model, we choose $\epsilon=10^{-6}$ and $m_{N0}=10^{10}~\mathrm{GeV}$. We show the Boltzmann evolution for the yield of the right-handed neutrino $Y_N$ (blue solid line) and the yield of the baryon asymmetry in Fig.~\ref{fig:evolutionDensity}. We compare the $Y_B$ evolution of the standard scenario without (green dashed line) and with our new contributions (orange solid line). The evolution in the weak washout (left panels) and strong washout (right panels) regime is shown for two different example values $g_L=g_Q= \{10^{-3}, 10^{-6}\}$. Generally, we observe that the equilibrium yield of the right-handed neutrino $Y_N^{\mathrm{(eq)}}$ is reached much faster in the strong washout regime due to the larger decay rate (cp. Fig.~\ref{fig:evolutionContribs}, green solid line). Additionally, we present the evolution of the different, relevant contributions in Fig.~\ref{fig:evolutionContribs}.\\


{\bf Scenario I ($\bm{g_L=g_Q=10^{-3}}$).} As naively expected, for relatively large couplings, the largest effect of the new TeV-scale LNV washout terms can be observed. Comparing Figs. \ref{fig:evolutionDensity_a} and \ref{fig:evolutionContribs_a} (weak washout), we see that the constant behavior of $Y_B$ at around $z\approx 0.6$ is caused by the dip that the decay rate $D$ receives due to the closing of $N\rightarrow HL$ and the opening of $H\rightarrow NL$ for small $z$. Even though the washout originating from $\widetilde{W}$ is stronger than the standard contribution $W$, it has no visible impact on the evolution of $Y_B$ (this picture changes for couplings of $\mathcal{O}(10^{-1})$). Around $z>7$, when the  $W$ contribution decreases strongly, the $\widetilde{W}$ contribution remains constant and leads to a strong washout such that $Y_B$ falls below the observed value for the baryon asymmetry in contrast to the standard leptogenesis scenario.\\

A comparable situation is found for the strong washout regime (Figs. \ref{fig:evolutionDensity_b} and \ref{fig:evolutionContribs_b}). Due to the larger coupling, $y_N$, the washout originating from the standard leptogenesis scenario $W$ is now dominant $z < 10$, see Fig.~\ref{fig:evolutionContribs_b}. Hence, the behavior of the standard leptogenesis scenario is followed longer up to larger $z$. However, when $W$ decreases significantly while $\widetilde{W}$ remains constant, $Y_B$ gets again fully washout out.\\

The $T$-independent washout term $\widetilde{W}$ for $z<10^6$ in Fig. \ref{fig:evolutionContribs} can be understood as follows. The dominant contribution to $\widetilde{W}$ is given by the inverse decay $\widetilde{D}$ involving $F\leftrightarrow SL$ or $S\leftrightarrow FL$. This expression is a function of the decaying particle's mass over temperature ($m_{F,S}/T$) and the right-handed neutrino mass ($m_N$). As shown in Fig. \ref{fig:thermalmasses}, for the relevant temperature range, both $m_F(T)$ and $m_S(T)$ are linear in temperature\footnote{The proportionality constant can be obtained from Eqs. \eqref{eq:mF(T)} and \eqref{eq:mS(T)}.} and $m_N(T)\approx m_{N0}$. Consequently, all quantities involved are effectively independent of the temperature. \\

The magnitude of $\widetilde{W}(z<10^6)$ can be naively estimated from a power counting on the lepton number violating couplings, generically referred to as $g$. Since this washout term is dominated by the inverse decay, and this process involves one lepton number violating vertex, then $\widetilde{W}=\mathcal{O}(g^2)$. It is important to highlight the fact that, in a zero temperature approximation, the most relevant contribution to $\widetilde{W}^{T=0}$ is given by the scattering terms. Since these terms involve two lepton number violating vertices, it follows that $\widetilde{W}^{T=0}=\mathcal{O}(g^4)$. Therefore, the inclusion of thermal effects is necessary to avoid an artificial suppression of the washout contribution coming from our simplified model. For lower temperatures, around $z=10^6$, the $\widetilde D$ contribution drops due to the mass hierarchy inversion, $ m_F(T)\approx m_S(T)$, leading to the corresponding drop in $\widetilde W$.\\

{\bf Scenario II ($\bm{g_L=g_Q=10^{-6}}$).} For smaller couplings, the washout contribution $\widetilde{W}$ is much smaller (see Figs. \ref{fig:evolutionContribs_c} and  \ref{fig:evolutionContribs_d}) and only at later times dominant in comparison to the conventional washout processes. Hence, the baryon asymmetry $Y_B$ drops below the observed baryon asymmetry also at later times for both the strong and weak washout scenario (see Figs. \ref{fig:evolutionDensity_c} and \ref{fig:evolutionDensity_d}).\\

\begin{figure}[t]
\centering
\includegraphics[width=\textwidth]{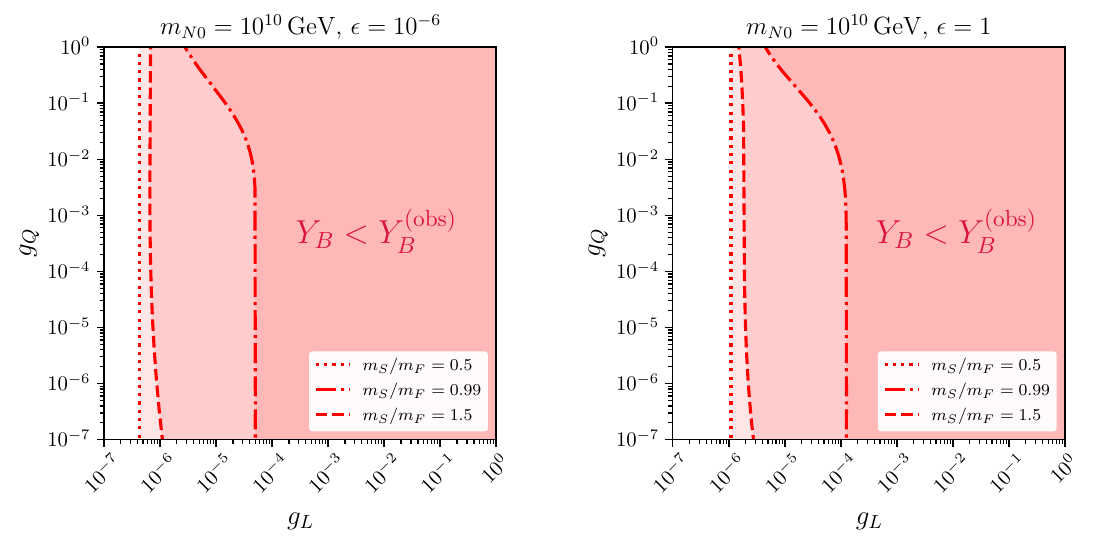}
\caption{$g_L$-$g_Q$--plane for $\epsilon=10^{-6}$ (left panel) and $\epsilon=1$ (right panel), with $m_{N0}=10^{10}~\mathrm{GeV}$ and $m_F=1~\mathrm{TeV}$. The red areas indicate the couplings that lead to a too-strong washout and don't result in the observed baryon asymmetry for the different mass hierarchies shown in the panels. Considering the weak ($K=10^{-2}$) or strong ($K=10^{2}$) washout regime does not affect the plot visibly. Notice how the extension of the viable thermal leptogenesis region depends on the new particle spectrum ---a feature not readily seen within the previously used pure EFT approach.}
\label{fig:LG_allowed_forbidden}
\end{figure}

Finally, we compare the final yield of the baryon asymmetry with the experimentally observed value in Eq.~\eqref{eq:BAUobs_Y}. The results are summarized in Fig.~\ref{fig:LG_allowed_forbidden}, choosing $m_{N0}=10^{10}~\mathrm{GeV}$. We show in red the parameter space that cannot account for the observed baryon asymmetry for the strong and weak washout scenario for two choices of the CP-asymmetry parameter, the usual choice $\epsilon=10^{-6}$ and the maximal possible CP-asymmetry $\epsilon=1$. We observe that standard thermal leptogenesis is rendered unviable for lepton-number violating couplings $g_L$ of the order of $10^{-6}$ for $m_S\neq m_F$ and $10^{-4}$ for $m_S\sim m_F$, respectively. It is important to highlight that the viable thermal leptogenesis region's extension relies heavily on the new particle spectrum, a characteristic not easily discernible when employing the previously used pure EFT approach. As previously noted, the predominant factor contributing to the TeV-scale washout, $\widetilde W$, is the inverse decay of the new particles $S$ or $F$. In a quasi-degenerate scenario, this decay is significantly suppressed. As a result, there is a more extensive viable region for leptogenesis compared to a hierarchical scenario. This demonstrates the value of our chosen methodology in revealing crucial information that might otherwise be obscured. The presence of a difference in two orders of magnitude significantly impacts the exploration of the viable region through long-lived particle searches. This substantial variation emphasizes the importance of accurately characterizing the particle spectrum, as it can greatly influence our understanding of the thermal history and collider phenomenology in leptogenesis studies. Moreover, a discovery of lepton-number violating new physics at collider or $\onbb$-decay experiments in this parameter range would have far-reaching consequences on the validity of standard thermal leptogenesis.\\

One can understand the weak dependence on $g_Q$ in Fig. \ref{fig:LG_allowed_forbidden} by observing that in the limit of $\lambda_{HS}\to0$ in Eq. \eqref{eq:ModelO2}, there is a complete correspondence between standard thermal leptogenesis with three heavy, RH Majorana neutrinos and the present simplified model scenario: $F$ and $S$ play the roles of the RH neutrino $N$ and the SM-Higgs $H$, respectively, and the couplings $g_Q$ and $g_L$ act as the quark Yukawa coupling and $y_N$, respectively. The TeV-scale washout $\widetilde W$ is dominated by the inverse decays, so as in some regimes of a standard thermal leptogenesis, the relevant parameters are $g_L \leftrightarrow y_N$. Moreover, in the limit of $g_Q\to0$, it is possible to assign a lepton number to $S$, thereby making the Lagrangian in Eq. \eqref{eq:ModelO2}  lepton number conserving. In this case, the asymmetry induced in the left-handed lepton sector is controlled only by $g_L$. As in Dirac leptogenesis \cite{Dick:1999je, Murayama:2002je}, this LH lepton number asymmetry can lead to non-vanishing $Y_B$, even though the total lepton number is conserved.

\section{Collider study}
\label{sec:collider}

The current experimental literature presents different results of various searches for LNV signals, ranging from specific decay modes of new particles (\emph{e.g.}, searches for $H^{\pm\pm}\to\ell^\pm\ell^\pm$ \cite{Chatrchyan:2012ya, Aaboud:2017qph}) to comprehensive studies of BSM theories (\emph{e.g.}, Left-Right symmetric models \cite{Aaboud:2018spl} or $R-$parity violating SUSY \cite{Chatrchyan:2013xsw}), including the connection with CP-violating effects at the LHC (\emph{e.g.}, rare $W^\pm$ decays \cite{Najafi:2020dkp}). The current status of those searches shows no evidence for significant deviations from the SM \cite{Khachatryan:2016kod, Sirunyan:2017uyt}.\\ 

\begin{figure}[h]
	\centering
	\includegraphics[width=0.7\textwidth]{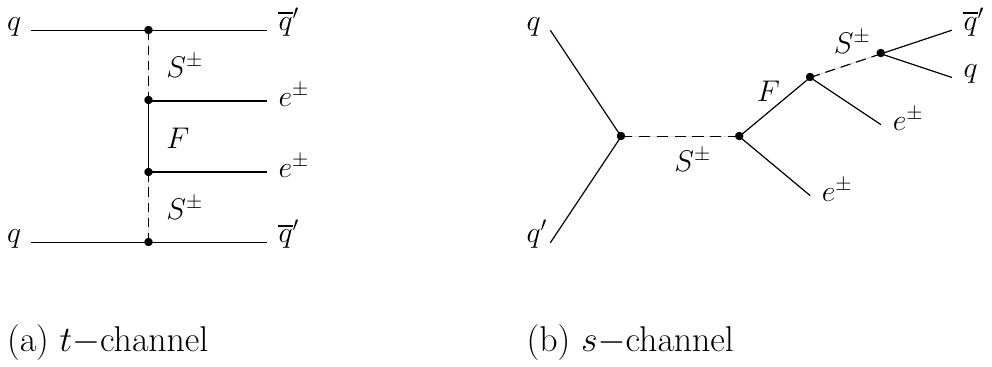}\\
	\caption{Feynman diagrams of the two-jet, same-sign dilepton signal ($pp\to e^\pm e^\pm jj$) in our simplified model. Diagram (a) matches the $\onbb-$decay diagram in Fig. \ref{fig:onbb_diagram}.}
	\label{fig:collider_diagrams}
\end{figure}

The goal of our work is to study the interplay and complementarity between collider phenomenology and $\onbb$-decay experiments. Since the latter one involves both electrons and quarks --at a fundamental level--, our analysis is focused on studying the production of two same-sign electrons and two jets in a proton-proton collider, namely $pp\to jje^\pm e^\pm$. Our simplified model allows two topologies associated with the signal, as shown in Fig. \ref{fig:collider_diagrams}.\\

In principle, some direct searches might also restrict our model. The first term in Eq. \ref{eq:ModelO2} makes this model sensitive to current experimental limits from di-jet resonant production. The ATLAS collaboration has recently published searches for low-mass \cite{Aaboud:2019zxd} and high-mass \cite{Aad:2019hjw} resonances in mass distributions of two jets. Reinterpreting those results, specifically the generic Gaussian-shaped distributions, we find that values for $g_Q\gtrsim0.1$ are roughly excluded for our parameter region of interest. The second term in Eq. \ref{eq:ModelO2} shows a potential sensitivity to single lepton plus missing $E_T$ searches \cite{CMS:2021pmn} if $F$ decays outside the detector. However, based on our estimations for the decay length \cite{Banerjee:2019ktv}, we conclude that $F$ will decay promptly for the parameter region of interest.\\

\subsection{Event generation and classification}
To perform our collider study, we have implemented the model \eqref{eq:ModelO2} using \textsc{FeynRules} \cite{Alloul:2013bka} to generate events with \textsc{Madgraph} \cite{Alwall:2014hca} at the parton level. We rely on \textsc{PYTHIA} \cite{Sjostrand:2014zea} for parton showering and jet matching, and \textsc{Delphes} \cite{deFavereau:2013fsa} for fast detector simulation. For both, signal and background, we impose a set of basic selection cuts ($p_{T_j},p_{T_\ell}>20\,\GeV$, $|\eta_j|<2.8$, $|\eta_\ell|<2.5$) at the generator level \cite{Peng:2015haa}, and a pre-selection rule ($N_j\geq2$, $N_{\ell^\pm\ell^\pm}\geq2$) at the classification level.\\

Our analysis is focused on the potential reach of the LHC at 14 TeV, in addition to the hypothetical FCC-hh \cite{Mangano:2017tke} and SppC \cite{Tang:2015qga} at 100 TeV. The \textsc{Delphes} software package \cite{deFavereau:2013fsa} incorporates the configuration card for ATLAS and CMS detectors at the LHC, while the equivalent one for the FCC-hh baseline detector is available online\footnote{FCC-hh detector \textsc{Delphes} card, \url{http://hep-fcc.github.io/FCCSW/}.} and a detailed description of the implementation is available in Ref. \cite{Selvaggi:2717698}.\\

The event classification is based on a custom-made Recurrent Neural Network (RNN) inspired by previous experiences \cite{Paganini:2157570} in the context of event discrimination using Deep Learning. An RNN allows us to classify events with unordered, variable length inputs, such as the number of jets or electrons \cite{Guest:2016iqz}. Our implementation uses the kinematic properties of jets, electrons, and missing $E_T$ to differentiate signal-like and background-like events. A detailed description of our topology is given in Appendix \ref{app_RNN} where we have also included a brief introduction to RNNs.\\

Additionally, we also implemented a Boosted Decision Tree (BDT) to compare these two machine learning approaches' results. We found that both the RNN and the BDT implementations presented a similar performance, consistent with previous comparisons available in the literature \cite{Bradshaw:2019ipy, Alves:2016htj, Buzatu:2651122}. As both implementations have similar performance, our choice of using an RNN is based on its ease of use since it offers a little more flexibility than a BDT.\\

\subsection{Signal generation and phenomenology}
The cross section associated with a $t-$channel production in Fig. \ref{fig:collider_diagrams}, $\sigma_{[t]}$, has a coupling dependence given by $(g_L\,g_Q)^4$ and it is relatively insensitive of the mass hierarchy between $F$ and $S^\pm$. Additionally, the $s-$channel cross section in Fig. \ref{fig:collider_diagrams}, $\sigma_{[s]}$, always involves the production of an on-shell particle. These two facts make the latest one to dominate the cross section over the first one, $\sigma(pp\to jje^\pm e^\pm)\approx\sigma_{[s]}(pp\to jje^\pm e^\pm)$. Consequently, the behavior of diagram (b) gives us insights about the total cross section.\\

To understand the coupling dependence of $\sigma_{[s]}$, notice that the physical processes vary depending on the mass hierarchy due to kinematic constraints as summarized in Table \ref{table:mass_hierarchy_sigma}. Each sub-process corresponds to the successive production or decay chains of the signal.\\

\begin{table}[ht]
	\centering
	\begin{tabular}{c@{\hskip 1cm}c@{\hskip 1cm}lll}
		\toprule
		{\bf Case} & {\bf Mass hierarchy} & \multicolumn{3}{c}{\bf Process} \\
		\midrule
		C1 & $m_S < m_F$ & $pp \to e^\pm F$, &$F \to e^\pm S^\mp$, & $S^\mp \to jj$  \\ 
		\midrule 
		C2 & $m_S = m_F$ & $pp \to e^\pm F$, &$F \to e^\pm jj$ &             \\ 
		\midrule
		C3 & $m_S > m_F$ & $pp \to S^\pm$,  &$S^\pm \to e^\pm F$, & $F \to e^\pm jj$ \\
		\bottomrule
	\end{tabular}
	\caption{Kinematic classification of production and successive decays involved in diagram (b), Fig. \ref{fig:collider_diagrams}, in our simplified model.}
	\label{table:mass_hierarchy_sigma}
\end{table}

If we use the narrow-width approximation, it is possible to decompose the different sub-processes in Table \ref{table:mass_hierarchy_sigma} in the following manner:

\begin{itemize}
	\item[C1.] When $F$ is heavier than $S^\pm$, the cross section corresponds to the production of an on-shell $F$ in addition to an electron --with $\sigma_{[s]}(pp\to e^\pm F)\propto (g_L\,g_Q)^2$-- followed by two cascade decays. These decay modes have branching ratios that are coupling independent since those are the only ones kinematically allowed:
	\begin{equation}
	\sigma_{[s]}(pp\to e^\pm e^\pm jj) = \underbrace{\sigma_{[s]}(pp\to e^\pm F)}_{\propto\ (g_L\,g_Q)^2}\times\underbrace{\mathrm{Br}(F\to e^\pm S^\mp)}_{=\ 2/3}\times\underbrace{\mathrm{Br}(S^\pm\to jj)}_{=\ 1}\,.
	\end{equation}
	
	\item[C2.] In the same fashion, if both $S^\pm$ and $F$ have equal masses then the cross section also corresponds to the production of an on-shell $F$ accompanied by an electron, followed by the decay of $F$ into a pair of jets and an electron. This decay, again, is the only one possible --so it has a branching ratio equal to one-- and it is mediated by an off-shell $S^\pm$ propagator:
	\begin{equation}
	\sigma_{[s]}(pp\to e^\pm e^\pm jj) = \underbrace{\sigma_{[s]}(pp\to e^\pm F)}_{\propto\ (g_L\,g_Q)^2}\times\underbrace{\mathrm{Br}(F\to e^\pm jj)}_{=\ 2/3}\,.
	\end{equation}
	
	\item[C3.] Finally, the case where $m_S > m_F$ provides a more subtle dependence. The full cross section can be thought as the on-shell production of $S^\pm$ --with $\sigma_{[s]}(pp\to S^\pm)\propto g_Q{}^2$-- followed by two successive decays. In this regime, $S^\pm$ is allowed to decay into two jets or a pair $e^\pm F$. The branching ratio is a function of the two couplings, as shown in Eq. \ref{eq:Br_S_Fe}:
	\begin{equation}
	\sigma_{[s]}(pp\to e^\pm e^\pm jj) = \underbrace{\sigma_{[s]}(pp\to S^\pm )}_{\propto\ g_Q{}^2}\times\underbrace{\mathrm{Br}(S^\pm \to e^\pm F)}_{=\ \left[1+\kappa\,\left(\frac{g_Q}{g_L}\right)^2\right]^{-1}}\times\underbrace{\mathrm{Br}(F\to e^\pm jj)}_{=\ 2/3}\,,
	\label{eq:Br_S_Fe}
	\end{equation}
	
	where $\kappa$ depends on the ratio between the two masses satisfying $\kappa>1$ for $m_S>m_F$. It is worthwhile to highlight two limiting cases:
	\begin{itemize}
		\item If $g_Q \gg g_L$, then $\mathrm{Br}(S^\pm \to e^\pm F) \propto (g_L/g_Q)^2$ and the cross section scales with $g_L{}^2$.
		\item If $g_Q \ll g_L$, then the $\mathrm{Br}(S^\pm \to e^\pm F)\approx 1$ and the cross section scales with $g_Q{}^2$.\\
	\end{itemize}
\end{itemize}

A key difference between the three cases is the magnitude of the respective cross sections. To illustrate, consider the production cross sections in cases C1-C2 and C3, \emph{i.e.}, $\sigma_{[s]}(pp\to e^\pm F)$ and $\sigma_{[s]}(pp\to S^\pm )$, respectively. In the first one, notice that the momentum transfer along the $s-$channel needed to produce an on-shell $F$ implies a suppression since the particle in the propagator is off-shell. However, the production in the latest one directly creates an on-shell $S^\pm$ so no suppression is applied. The different order in the couplings reinforces the difference between the magnitudes of the cross sections for the same set of parameters, as we illustrate in Table \ref{table:example_sigma_mSmF}.\\

\begin{table}[ht]
	\centering
	\begin{tabular}{l@{\hskip 2.5cm}c@{\hskip 1cm}c}
		\toprule
		& $m_S=0.5\,\TeV$  & $m_S=2\,\TeV$ \\ \midrule
		\multicolumn{1}{l@{\hskip 2.5cm}}{$\sigma(pp\to jje^\pm e^\pm)$} & $1.199\times10^{-8}\,\mathrm{pb}$    & $8.874\times10^{-5}\,\mathrm{pb}$ \\
		\multicolumn{1}{l}{$\sigma_{[s]}(pp\to jje^\pm e^\pm)$}                & $1.195\times10^{-8}\,\mathrm{pb}$     & $8.874\times10^{-5}\,\mathrm{pb}$ \\ \midrule
		\multicolumn{1}{l}{Production mode}                   & $pp\to e^\pm F$                                         & $pp\to S^\pm$ \\
		\multicolumn{1}{l}{Production cross section}     & $4.012\times10^{-8}\,\mathrm{pb}$       & $3.545\times10^{-4}\,\mathrm{pb}$                \\
		\bottomrule
	\end{tabular}
\caption{Numerical example of the phenomenological behavior of the signal cross section at $\sqrt{s}=14\,\TeV$. We took $m_F=1\,\TeV$, $g_L=0.1$, and $g_Q=0.01$ in both scenarios. The results were obtained from \textsc{Madgraph} \cite{Alwall:2014hca}.}
\label{table:example_sigma_mSmF}
\end{table}

\subsection{Background generation and validation}
Backgrounds in the same-sign dilepton final state can be divided into three categories \cite{Khachatryan:2016kod}:

\begin{itemize}
	\item {SM processes with same-sign dileptons}, including diboson production (considering $W$, $Z$, $H$, and prompt $\gamma$), single boson production in association with a $t\overline{t}$ pair, and ``rare'' processes (\emph{e.g.}, $t\overline{t}t\overline{t}$ and double-parton scattering).
	
	\item {Charge misidentification} from events with opposite-sign isolated leptons in which the charge of an electron is misidentified, mostly due to severe \textit{bremsstrahlung} in the tracker material.
	
	\item {Jet-fake leptons} from heavy-flavor decays, where hadrons are misidentified as leptons, or electrons from unidentified conversions of photons in jets.
\end{itemize}

For our analysis, we study the effects of the dominant contributions: EW diboson processes, charge misidentification involving $\gamma/Z^\ast$, and jet-fakes produced by $t\overline{t}$ and $W+3j$ processes \cite{Peng:2015haa, Sirunyan:2017uyt}. The diboson simulation involved the generation of $WW$, $WZ$, and $ZZ$ events plus jets, and their respective leptonic decays, as implemented in Ref. \cite{Peng:2015haa}.\\

Due to the difficulty of precisely simulating jet-fakes, we implement the ``FakeSim'' method proposed in Ref. \cite{Curtin:2013zua} as an additional module in \textsc{Delphes}. This data-driven approach takes into account the relation between the originating jet and the fake lepton. It is based on two functions\footnote{In the FakeSim method, these functions are parameterized by four quantities, namely $\{\epsilon_{200},r_{10},\mu,\sigma\}$. See Ref. \cite{Curtin:2013zua} for additional details.\label{footnote_fakesim}}:
\begin{enumerate}
	\item A mistag efficiency, $\epsilon_{j\to\ell}(p_{T_j})$, representing the probability that a particular jet $j$ is mistagged as a lepton $\ell$.
	\item A transfer function, $\mathcal{T}_{j\to\ell}$, modeling the probability distribution function that maps $p_{T_j}$ into the fake $p_{T_\ell}$.
\end{enumerate}

\begin{figure}[ht]
	\centering
	\begin{minipage}{0.5\textwidth}
		\centering
		\includegraphics[width=0.9\textwidth]{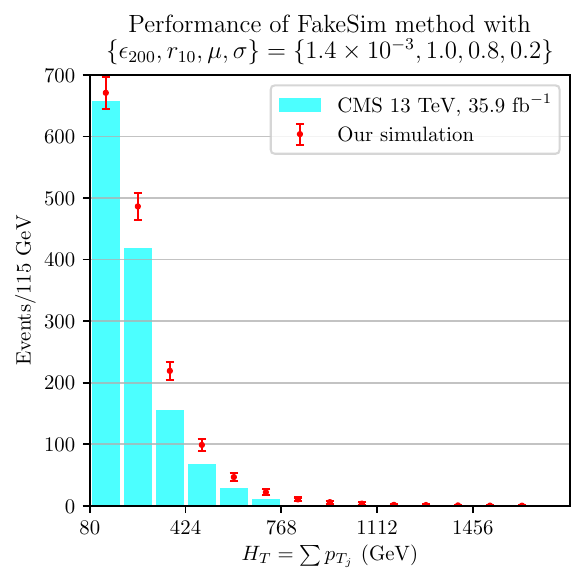}\\
	\end{minipage}%
	\begin{minipage}{0.5\textwidth}
		\centering
		\includegraphics[width=0.9\textwidth]{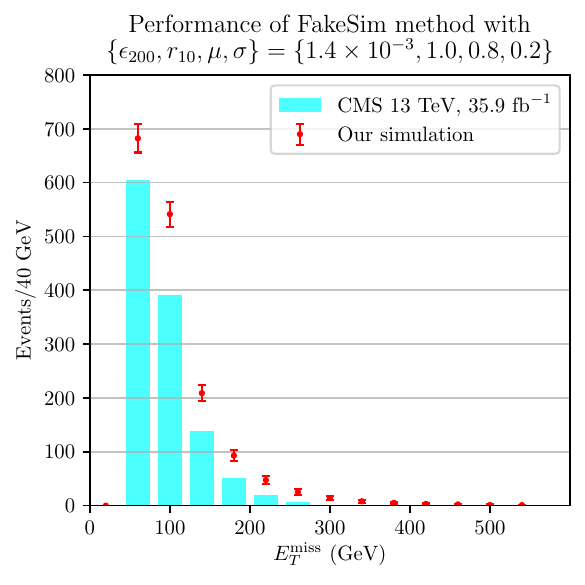}\\
	\end{minipage}
	\caption{Comparison between the CMS results in Ref. \cite{Sirunyan:2017uyt} and our implementation of the FakeSim method proposed in Ref. \cite{Curtin:2013zua} for a particular choice of its parameters, as mentioned in Note \ref{footnote_fakesim}. Distributions of the kinematic variables $H_T$ and $E_T^\mathrm{miss}$ are shown in the left and right panels, respectively.}
	\label{fig:fakesim_validation}
\end{figure}

Using data or simulations from ATLAS or CMS collaborations, it is possible to fit the set of parameters to find consistent results for different phenomenological studies \cite{Izaguirre:2015pga, Dib:2016wge, Nemevsek:2016enw}. In Fig. \ref{fig:fakesim_validation}, we compare representative CMS results (digitized from Fig. 3 in Ref. \cite{Sirunyan:2017uyt}) with our result obtained using the FakeSim method. As can be seen from the plot, we reproduce the overall behavior after a parameter fitting. It is possible to optimize the choice of the FakeSim parameters if, for instance, flavor effects are included by introducing flavor-dependent mistag efficiencies \cite{Priv_Comm_Li}.\\

To estimate the charge misidentification background, we introduce a misidentification probability for electrons from current doubly charged Higgs boson searches by the ATLAS collaboration \cite{Aaboud:2017qph}. The electron charge misidentification probability is modeled as a separable function of electron’s $p_T$ and $\eta$, $P(p_T,\eta)=\sigma(p_T)f(\eta)$, and its data-driven values are extracted from Fig. 3 in Ref. \cite{Aaboud:2017qph}. We follow a common approach in collider phenomenological studies \cite{Helo:2018rll, Pan:2019wwv} to incorporate this probability as a weight in opposite-sign generated events, as detailed in Section 7.1.3 in Ref. \cite{Muskinja:2643902}. In Table \ref{table:charge_misID}, we compare our background estimation with the ATLAS results extracted from Fig. 2 in Ref. \cite{Aaboud:2017qph} for a $Z\to ee$ peak data sample. The ratio of same-charge/opposite-charge events for ATLAS is 0.93\% and we obtained 0.70\%.\\

\begin{table}[ht]
	\centering
	\begin{tabular}{l@{\hskip 1cm}c@{\hskip 1cm}c@{\hskip 1cm}c}
		\toprule
		& \multicolumn{3}{@{\hskip 1cm}c@{\hskip 1cm}}{\bf Number of events} \\ \cmidrule{2-4} 
		& Opposite-charge & Same-charge & SC/OC \\
		& (OC) & (SC) & ratio \\\midrule
		ATLAS & $1.23\times10^7$ & $1.14\times10^5$ & 0.93\% \\ \midrule
		Our estimation & $9.41\times10^6$     & $6.59\times10^4$ & 0.70\% \\
		\bottomrule
	\end{tabular}
	\caption{Validation of the charge misidentification implementation. The ATLAS results extracted from Fig. 2 in Ref. \cite{Aaboud:2017qph}, and our results make use of the probability density defined in Ref. \cite{Muskinja:2643902}.}
	\label{table:charge_misID}
\end{table}

Notice that the two data-driven methods previously described were validated using LHC data. Table \ref{table:bkg_lhc} presents the background cross section for the three classes, detailing the effects of the signal selection rule $(N_j\geq 2,\ N_{\ell^\pm \ell^\pm}\geq2)$ and the background discrimination by using the RNN. The first column shows the cross section before the signal selection ($\sigma_\mathrm{before}$), as given directly by \textsc{Madgraph}. The second column shows the cross sector after applying the signal selection rule ($\sigma_\mathrm{after}$), where the number of events is initially reduced. Finally, the third column shows the background cross section classified by the RNN ($\sigma_\mathrm{RNN}$), and we notice the background reduction provided by our machine learning implementation. Since there are no estimations for these types of background for a future 100 TeV hadron collider, we use the same set of parameters and functions varying only the energy of the center of mass. Table \ref{table:bkg_fcc} presents the results of our estimation.\\

\begin{table}[ht]
	\centering
	\begin{tabular}{cc@{\hskip 1cm}c@{\hskip 1cm}c@{\hskip 1cm}c}
		\toprule
		\multicolumn{2}{c}{\bf Background type} & $\sigma_\mathrm{before}$ (pb) & $\sigma_\mathrm{after}$ (pb) & $\sigma_\mathrm{RNN}$ (pb) \\
		\midrule
		\multirow{3}{*}[-4.5pt]{Diboson} & $WW$ & $3.28\times10^{-3}$ & $6.40\times10^{-4}$ & $6.87\times10^{-5}$ \\ \cmidrule{2-5} 
		& $WZ$ & $2.59\times10^{-2}$ & $6.65\times10^{-3}$ & $2.10\times10^{-4}$ \\ \cmidrule{2-5} 
		& $ZZ$ & $1.32\times10^{-3}$ & $5.62\times10^{-4}$ & $1.14\times10^{-5}$ \\
		\midrule
		\multirow{2}{*}[-2pt]{Jet-fake} & $W+3j$ & $1.79\times10^{-1}$ & $4.34\times10^{-2}$ & $1.78\times10^{-4}$ \\ \cmidrule{2-5} 
		& $t\bar{t}$ & $9.11\times10^{-2}$ & $2.64\times10^{-2}$ & $6.10\times10^{-5}$ \\
		\midrule
		\multirow{2}{*}[-2pt]{Charge misidentification} & $t\bar{t}$ & $3.33\times10^{-2}$ & $1.54\times10^{-2}$ & $4.45\times10^{-4}$ \\ \cmidrule{2-5} 
		& $Z/\gamma^\ast$ & $2.54\times10^{-1}$ & $1.37\times10^{-1}$ & $4.89\times10^{-3}$ \\
		\midrule
		& & $5.88\times10^{-1}$ & $2.30\times10^{-1}$ & $5.86\times10^{-3}$\\
		\bottomrule
	\end{tabular}
	\caption{Individual contributions to the cross section for the different types of background for the LHC at 14 TeV. $\sigma_\mathrm{before}$ corresponds to the cross section before the signal selection, $\sigma_\mathrm{after}$ corresponds to the cross section after the signal selection, and $\sigma_\mathrm{RNN}$ is the cross section tagged as background events by the RNN. The last row shows the total background cross section.}
	\label{table:bkg_lhc}
\end{table}

\begin{table}[ht]
	\centering
	\begin{tabular}{cc@{\hskip 1cm}c@{\hskip 1cm}c@{\hskip 1cm}c}
		\toprule
		\multicolumn{2}{c}{\bf Background type} & $\sigma_\mathrm{before}$ (pb) & $\sigma_\mathrm{after}$ (pb) & $\sigma_\mathrm{RNN}$ (pb) \\
		\midrule
		\multirow{3}{*}[-4.5pt]{Diboson} & $WW$ & $3.56\times10^{-2}$ & $1.76\times10^{-2}$ & $2.27\times10^{-3}$ \\ \cmidrule{2-5} 
		& $WZ$ & $3.78\times10^{-1}$ & $1.88\times10^{-1}$ & $5.88\times10^{-3}$ \\ \cmidrule{2-5} 
		& $ZZ$ & $1.95\times10^{-2}$ & $1.35\times10^{-2}$ & $2.53\times10^{-4}$ \\
		\midrule
		\multirow{2}{*}[-2pt]{Jet-fake} & $W+3j$ & $2.21$ & $7.16\times10^{-1}$ & $6.72\times10^{-3}$ \\ \cmidrule{2-5} 
		& $t\bar{t}$ & $3.85$ & $1.43$ & $6.82\times10^{-3}$ \\
		\midrule
		\multirow{2}{*}[-2pt]{Charge misidentification} & $t\bar{t}$ & $5.11\times10^{-1}$ & $3.36$ & $1.60\times10^{-1}$ \\ \cmidrule{2-5} 
		& $Z/\gamma^\ast$ & $3.33$ & $2.02$ & $1.51\times10^{-1}$ \\
		\midrule
		& & $10.33$ & $7.75$ & $3.33\times10^{-1}$\\
		\bottomrule
	\end{tabular}
	\caption{Individual contributions to the cross section for the different types of background for the FCC-hh at 100 TeV. The definitions of $\sigma_\mathrm{before}$, $\sigma_\mathrm{after}$, and $\sigma_\mathrm{RNN}$ are equivalent as in Table \ref{table:bkg_lhc}. The last row shows the total background cross section.}
	\label{table:bkg_fcc}
\end{table}

\section{\texorpdfstring{\boldmath{$\onbb$}}{0nbb} Decay}
\label{sec:nldbd}

The results from searches for $\onbb$ place complementary constraints on the model parameters in a manner that can complement the collider search results. To date, they have yielded null results, with the present strongest limit on the half-life of $^{136}$Xe having been set by the KamLAND-Zen experiment \cite{KamLAND-Zen:2016pfg}\\
\begin{equation}
T^{0\nu}_{1/2} > 2.3 \times 10^{26}~\textrm{years at 90\% C.L.}\,
\end{equation}

The next generation of \lq\lq tonne-scale" $0\nu\beta\beta$ decay searches aim to increase this sensitivity by two orders of magnitude, with a variety of isotopes under consideration \cite{Albert:2017hjq, Kharusi:2018eqi, Abgrall:2017syy, Armengaud:2019loe, CUPIDInterestGroup:2019inu, Paton:2019kgy}.\\

The $\onbb$ decay mechanism is a combination of two contributions, one from the dimension-5 Weinberg operator (or equivalently, the Majorana mass term for neutrinos), and the dimension-9 operator $\mathcal{O}_1$ in Eq. \eqref{eq:eq3}. From simple dimensional analysis, one finds that the ratio of the amplitudes for the two contributions is\footnote{A similar comparison is done in Ref. \cite{Deppisch:2015yqa} by setting the scale of the dimension-5 amplitude of the effective Majorana mass to the upper experimental limit.}

\begin{equation}
\frac{\mathcal{A}_{d=9}}{\mathcal{A}_{d=5}} = \frac{\langle p\rangle^2\,C_1}{G_F^2\,m_{\beta\beta}\,\Lambda^5} \,,
\end{equation}

where $m_{\beta\beta}$ is the effective neutrino mass entering the light neutrino amplitude and $\langle p \rangle$ is the light neutrino virtuality. For typical values of $m_{\beta\beta}$ and $\langle p \rangle$, these two contributions are comparable when $C_1\sim\mathcal{O}(1)$ and $\Lambda\sim\mathcal{O}(1\,{\rm TeV}).$\\

The interpretation of a $\onbb$ decay result thus depends on the presence or absence of $\mathcal{A}_{d=9}$ \cite{Cirigliano:2004tc, Deppisch:2017ecm}. It is possible that the effect of $\mathcal{A}_{d=9}$ can be canceled against the $\mathcal{A}_{d=5}$ contribution. Pessimistically, even if neutrinos exhibit the inverse hierarchy (IH), one may not observe a signal in a tonne-scale experiment due to this cancellation, which would suppress the rate. On the other hand, additional BSM physics may enhance the signal. It may be, in such cases, that the light neutrinos exhibit a normal hierarchy (NH) with $\mathcal{A}_{d=5}$ well below the observable scale, yet a non-zero $\onbb$-decay signal arises to the presence of $\mathcal{A}_{d=9}$. 
In any of these scenarios, the experimental $\onbb$-decay result will have far-reaching consequences, and its theoretical interpretation -- including the possible implications for the viability of standard thermal leptogenesis -- will require additional information.\\

Motivated by the complementarity between collider searches and $\onbb$ decay experiments, we study a scenario where the dimension-9 contribution dominates the $\onbb$ decay rate.
To account for the different scales, we need to evolve the operator $\mathcal{O}_1$ in Eq.~\eqref{eq:eq3} from the TeV scale to the GeV scale using the renormalization group running. Hereby, operators receive QCD and electroweak corrections and mix with other operators. The RGE evolution was studied in Ref. \cite{Peng:2015haa}, with the dominant QCD corrections considered. The relevant part of $\mathcal{O}_1$ contributing to $\onbb$-decay is:

\begin{equation}
\mathscr{L}_\mathit{LNV}^\mathrm{eff} =\frac{C_\mathrm{eff}}{2\Lambda^5}\left(\mathcal{O}_{2+}^{++} - \mathcal{O}_{2-}^{++}\right)\overline{e_L}e_R^c\ + \ \mathrm{h.c.}\,,
\label{eq:O2eff}
\end{equation}

where $e_R^c\equiv(e_L)^c$, $C_\mathrm{eff} \approx C_1(1\,\mathrm{GeV}) = 0.092C_1(\Lambda)$~\cite{Peng:2015haa}, and the operator $\mathcal{O}_2$ being \cite{Prezeau:2003xn}: 
 
\begin{equation}
\mathcal{O}_{2\pm}^{ab} =(\bar{q}_R\tau^aQ)(\bar{q}_R\tau^bQ)\pm(\bar{Q}\tau^aq_R)(\bar{Q}\tau^bq_R)\,.
\label{eq:O2_L}
\end{equation}

\begin{figure}
	\centering
	\includegraphics[scale=0.8]{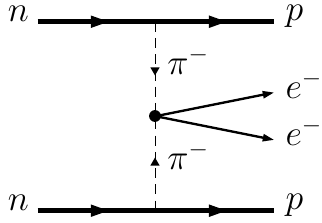} \\
	\caption{Contribution of dim-9 LNV $\pi\pi ee$ interaction to $\onbb-$decay at tree level.}
	\label{fig:pipiee}
\end{figure}

When $a = b$, the operator with subscript $+ (-)$ are even (odd) eigenstates of parity. As the hadronic (four quark) part of the operator carries no dependence on the lepton kinematics, the matrix element for the process factorizes into a leptonic and hadronic part. Computation of the former is straightforward. For the latter, we first match the four-quark operator $\mathcal{O}_{2+}^{++}-\mathcal{O}_{2-}^{++}$ onto hadronic degrees of freedom most appropriate for computation of the nuclear transition matrix element, following the effective field theory (EFT) approach delineated in Refs.~\cite{Prezeau:2003xn, Cirigliano:2018yza}. The leading order contribution to the nuclear matrix element (NME) arises from the pion-exchange amplitude of Fig.~\ref{fig:pipiee}, where the LNV $\pi\pi ee$ interaction emerges from matching $\mathcal{O}_{2+}^{++} {\bar e_L} e_R^c$ onto the two pion-two electron operator in Eq. \eqref{eq:O2eff}:

\be
\label{eq:pioneff}
\frac{C_\mathrm{eff}\Lambda_H^2 F_\pi^2}{2\Lambda^5} \pi^-\pi^- {\bar e_L} e_R^c +\mathrm{h.c.}\, ,
\ee

where $F_\pi=92.2\pm 0.2$ MeV  is the pion decay constant \cite{DGH:14} and $\Lambda_H$ is a mass scale associated with the hadronic matrix element (HME) of the four quark operator $\mathcal{O}_{2+}^{++}$,
\be
\label{eq:LamHdef}
\Lambda_H^2 F_\pi^2 =\frac{1}{2}\bra{0} \left(\mathcal{O}_{2+}^{++} -\mathcal{O}_{2-}^{++}\right)\ket {\pi^- \pi^-}\,.
\ee
 In the earlier work of Ref.~\cite{Peng:2015haa} $\Lambda_H^2$ was obtained using factorization/vacuum saturation to estimate the HME, yielding  $ \Lambda_H^2 = m_\pi^4/(m_u+m_d)^2 \approx -7.5$ GeV$^2$ for $m_{\pi^+}= 139$ MeV and $m_u+m_d=7$ MeV. Subsequently, the authors of Ref. \cite{Cirigliano:2017ymo} noted that one may relate $\mathcal{O}_{2+}^{++}$ to analogous $\Delta S\not=0$ four-quark operators using SU(3) flavor symmetry. Consequently, one may exploit flavor SU(3) to obtain estimates of $\Lambda_H^2$  from the corresponding strangeness changing $K^0\to{\bar K^0}$ and $K\to\pi\pi$ matrix elements. The result yields  $\Lambda_H^2 = -(3.16\pm 0.7)$ GeV$^2$ at the matching scale $\mu=3$ GeV. The Cal-Lat collaboration performed a direct computation of the matrix element in Eq.~(\ref{eq:LamHdef}), obtaining $\Lambda_H^2 = -(2.15\pm 0.36)$ GeV$^2$ at $\mu=3$ GeV in the RI/SMOM scheme \cite{Nicholson:2018mwc}. In what follows, we will adopt the Cal-Lat value.\\ 
 
 When used to evaluate the amplitude in Fig. \ref{fig:pipiee}, the interaction in Eq.~(\ref{eq:pioneff}) yields an effectivce two nucleon-two electron operator whose nuclear matrix elements (NMEs) may be evaluated using state-of-the-art many-body methods. The resulting expression for the decay rate is
 \bea
\label{eq:rate}
\frac{1}{T_{1/2}} & = & \left[G_{0\nu}\times (1\, \mathrm{TeV})^2\right] \left(\frac{\Lambda_H}{\mathrm{TeV}}\right)^4 \left(\frac{1}{144}\right) \\
&\times&
\nonumber
\left(\frac{v}{\mathrm{TeV}}\right)^8
\left(\frac{1}{\cos\theta_C}\right)^4 \vert M_0\vert^2 \, \left[\frac{C_\mathrm{eff}^2}{(\Lambda/\mathrm{TeV})^{10}}\right]\ \ \ , \\
\nonumber
G_{0\nu} &=& (G_F \cos\theta_C g_A)^4 \left(\frac{\hbar c}{R}\right)^2 \left(\frac{1}{32\pi^2 \hbar\ln 2}\right) I(E_{\beta\beta})\ \ \ ,
\eea
with $\theta_C$ being the Cabibbo angle, $I(E_{\beta\beta})$ the electron phase space integral
\be
 \int_{m_e}^{E_{\beta\beta}-m_e}  dE_1\, F(Z+2, E_1) F(Z+2, E_2) p_1 E_1 p_2 E_2\ \ \ , 
\ee
 $E_2=E_{\beta\beta} - E_1$, and $F(Z+2, E_{1,2})$ being factors that account for distortion of the electron wave functions in the field of the final state nucleus. 
The NME is given by 
\be
\label{eq:nme}
M_0 = \left\langle{\Psi_f} \right\vert \sum_{i\not=j} \frac{R}{\rho_{ij}} \left[ F_1 {\vec\sigma}_i\cdot {\vec\sigma}_j + F_2 T_{ij}\right]\tau^+_i \tau^+_j\left\vert {\Psi_i}\right\rangle \,,
\ee
where $T_{ij} = 3{\vec\sigma}_i\cdot{\hat\rho}_{ij} {\vec\sigma}_j\cdot{\hat\rho}_{ij} -{\vec\sigma}_i\cdot {\vec\sigma}_j$, $R=r_0 A^{1/3}$,  ${\vec\rho}_{ij}$ is the separation between nucleons $i$ and $j$, and the functions $F_{1,2}(|\vec\rho_{ij} |)$ are given in Ref.~\cite{Prezeau:2003xn}. Note that we have normalized the rate to the conventionally-used factor $G_{0\nu}$ that contains quantities associated with the SM weak interaction, even though the LNV mechanism here involves no SM gauge bosons. Note also that Eq.~(\ref{eq:rate}) corrects two errors in the corresponding expression in Ref.~\cite{Peng:2015haa}: (a) the inclusion here of a factor of $g_A^4$ and (b) an additional overall factor of $1/8$. The latter arises from a factor of $1/2$ due to the presence of two electrons in the final state and a factor of $(1/2)^2$ that one must include to avoid double counting in the NME since the sum runs over $i\not= j$ rather than $i<j$.\\

In the analysis of Ref.~\cite{Peng:2015haa}, a value of $M_0=-1.99$ for the transition ${^{76}\mathrm{Ge}}\to {^{76}\mathrm{Se}}$ was adopted from the quasiparticle random phase approximation (QRPA) computation of Ref.~\cite{Faessler:1998qv}. Here, we use the results of a more recent proton-neutron (pn) QRPA computation of Ref.~\cite{Hyvarinen:2015bda}. The resulting value for $M_0({^{76}\mathrm{Ge}}\to {^{76}\mathrm{Se}}) = -4.74$. At present, the most sensitive limit on the half life has been obtained using $^{136}$Xe, for which the matrix element in Ref.~\cite{Hyvarinen:2015bda} is $M_0({^{136}\mathrm{Xe}}\to {^{136}\mathrm{Ba}}) = -2.63$. Both pnQRPA values assume no \lq\lq quenching\rq\rq\ of $g_A$.\\

It is important to emphasize that the calculated NMEs exhibit considerable theoretical uncertainties. The earlier work of Ref.~\cite{Peng:2015haa} accounted for the combined effect of these uncertainties as well as those in HMEs by varying the value of $M_0$ by a factor of two. The subsequent chiral SU(3) and lattice computations of $\bra{0} \mathcal{O}_{2+}^{++} \ket {\pi^- \pi^-}$ have reduced the hadronic uncertainty to the $\mathcal{O}(10\%)$ level. In the case of the NME, however, it has been realized that in the context of few-nucleon effective field theory, consistent renormalization requires the presence of a contact interaction in addition to the long-range two-pion exchange amplitude \cite{Cirigliano:2018hja}. The corresponding operator coefficient and nuclear matrix element are presently unknown. We thus retain a factor of two uncertainty in the NME to account for both the {\em bona fide} nuclear many-body uncertainties as well as the effect of the \lq\lq counterterm\rq\rq\ contribution.

\section{Combined results and discussion}
\label{sec:discussion}
\begin{figure}[t]
	\centering
	\includegraphics[width=\textwidth]{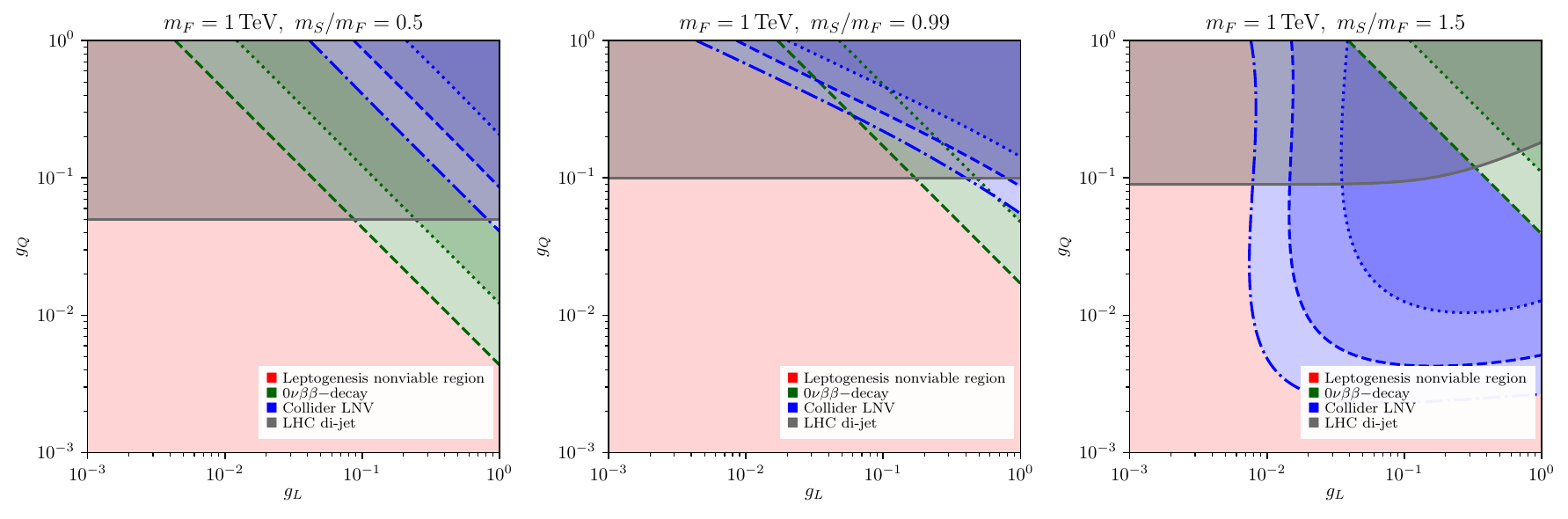}
	\caption{Interplay between leptogenesis, collider searches, and $\onbb-$decay experiments for $m_F=1$ TeV, and $m_S/m_F=\{0.5,0.99,1.5\}$. As in Fig. \ref{fig:LG_allowed_forbidden}, we present the nonviable leptogenesis region in red; in dark green, we show the $\onbb-$decay exclusion from KamLAND-Zen (dotted line) and future tonne-scale experiments (dashed line); the collider LNV (same sign dilepton plus dijet) exclusion is shown in blue for the LHC at 14 TeV with integrated luminosities of 100 fb$^{-1}$ (dotted line) and 3 ab$^{-1}$ (dashed line), and the FCC-hh at 100 TeV with 30 ab$^{-1}$ (dash-dotted line); and, in dark gray, we present the LHC di-jet exclusion, as discussed in Section \ref{sec:collider}.}
	\label{fig:scan_exclusion}
\end{figure}
In the following, we combine our results from the previous sections in order to investigate the reach and interplay of collider and $\onbb$-decay experiments and the implications of a possible discovery for the generation of the baryon asymmetry.\\

In Fig.~\ref{fig:LG_allowed_forbidden}, we illustrated the parameter space (red) that was found to lead to a baryon asymmetry smaller than the observed one because of a washout too strong due to the new TeV-scale interactions. Fixing the masses of the new particles $F$ and $S$ around the TeV scale, we can read off the couplings $g_L$ and $g_Q$, for which we would expect a strong washout that would prevent a large enough baryon asymmetry. Hence, a discovery of a process within this parameter space would preclude the viability of the standard thermal leptogenesis scenario. Notice that our findings are constrained by the assumption of the new particle $F$ to be in equilibrium. In that sense, the boundaries in the parameter space shown in Fig. \ref{fig:LG_allowed_forbidden} are subject to uncertainty from the calculations involving the abundance of $F$.\\

As depicted in Fig.~\ref{fig:onbb_diagram}, the new interactions in the Lagrangian, Eq.~\eqref{eq:ModelO2}, can lead to $\onbb$-decay. We show in green in Fig.~\ref{fig:scan_exclusion}, the region currently excluded region by KamLAND-Zen (green dotted line), but also the future reach of tonne-scale experiments (green dashed line). With this region lying in the red area, we can conclude that an observation of $\onbb$-decay realized by a dim-9 operator with new physics at the TeV scale would rule out the standard leptogenesis paradigm. As demonstrated in Fig.~\ref{fig:LG_allowed_forbidden} (right), this conclusion remains valid even for the most optimistic choice of maximal CP asymmetry.\\

With $\onbb$-decay being a low energy process, it is not sensitive to the mass hierarchy $m_S/m_F$ and explicit couplings $g_L$ and $g_Q$ of the new degrees of freedom at the TeV scale, but only to the effective coupling $C_1$ and scale $\Lambda$, as stated in Eq.~\eqref{eq:matching}. Therefore, the decreased reach from left to right in Fig.~\ref{fig:scan_exclusion}, is caused by the increase in $\Lambda$, leading to a suppression of the process.\\

High-energy collider experiments, however, can resolve the TeV-mass scale, and hence depend on the mass hierarchy of the new particles $F$ and $S$. We show the current 14 TeV collider limits with $100~\mathrm{fb}^{-1}$ integrated luminosity for the different mass hierachies (blue dotted line), as well as for future $3~\mathrm{ab}^{-1}$ integrated luminosity (blue dashed line) and the FCC-hh with 100 TeV and $30~\mathrm{ab}^{-1}$ (blue dashed-dotted line). As discussed in detail in Sec.~\ref{sec:collider}, the mass hierarchy is crucial for the reach of the collider searches. For $m_S>m_F$, $S$ can be produced resonantly followed by subsequent decays into a signature of two same-sign electrons and two jets, leading to the strongest constraints (see the dominant s-channel process depicted in Fig.~\ref{fig:collider_diagrams} (b)). We also reiterate that for $g_Q > g_L$ the limits are mainly restricted by $g^2_L$ and independent of $g_Q$, while for $g_L > g_Q$ the limits are mainly constrained by $g^2_Q$ (and not $g_L$). For $m_S \lesssim m_F$, in contrast, the collider constraints are much weaker, as $S$ can be produced only off-shell. If we compare the the current and future collider sensitivities (blue) to the red region for all three mass hierarchies, we see that an observation of two same-sign electrons and two jets would similarly rule out the standard thermal leptogenesis scenario for weak and strong washout.\\

While a same sign di-electron plus di-jet signature directly points to a LNV process, direct searches for the particles $F$ and $S$ also constrain the model parameter space. We show the limits from dijet resonant searches in gray. While covering already the full collider reach for $m_S \lesssim m_F$, they are less sensitive for $m_S > m_F$. The interplay of the di-jet, same sign di-electron, and $0\nu\beta\beta$-decay sensitivities in Fig. \ref{fig:scan_exclusion} illustrates the complementarity of these probes. Drawing on all of them will be important if a non-zero LNV signal is seen at either low- or high-energies.\\

As apparent from Fig.~\ref{fig:scan_exclusion}, the relative reaches of $\onbb$-decay experiments and collider LNV searches depend decisively on the new particle spectrum. This emphasizes the importance of pursuing and combining the low- and high-energy frontier in order to cover wide ranges of the parameter space but also to identify, in case of an observation, the underlying new physics. For instance, for $m_S \gtrsim m_F$, the observation of $\onbb$-decay would imply an observable signal at the LHC or would point towards another underlying $0\nu\beta\beta$-decay mechanism.\\

\begin{figure}[t]
	\centering
	\includegraphics[width=\textwidth]{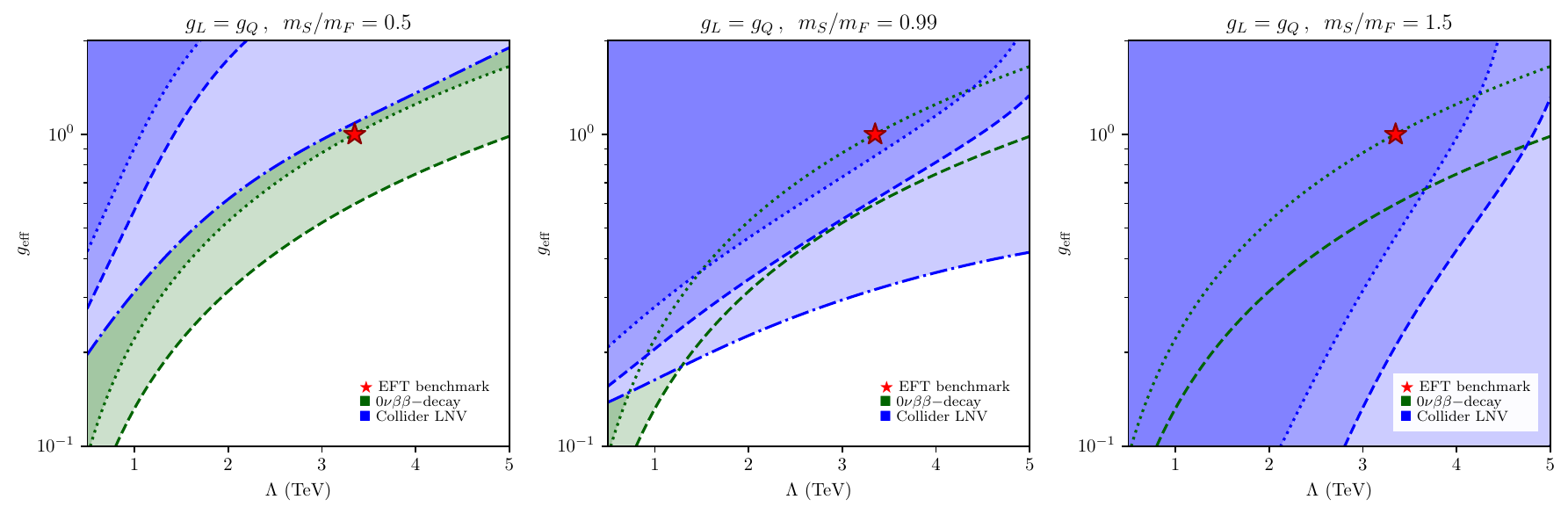}\\
	\caption{
		Complementarity between collider searches and $\onbb-$decay experiments for $g_L=g_Q$ and $m_S/m_F=\{0.5,0.99,1.5\}$. Comparison with an EFT analysis for the scale of new physics $\Lambda=(m_S^4m_F)^{1/5}$ and the effective coupling $g_\mathrm{eff}=g_L=g_Q$. In dark green, we show the $\onbb-$decay exclusion from KamLAND-Zen (dotted line) and future tonne-scale experiments (dashed line); and the collider LNV exclusion is shown in blue for the LHC at 14 TeV with integrated luminosities of 100 fb$^{-1}$ (dotted line) and 3 ab$^{-1}$ (dashed line), and the FCC-hh at 100 TeV with 30 ab$^{-1}$ (dash-dotted line). The EFT benchmark point is shown as a red star, where $g_\mathrm{eff}=1$ is assumed for the discussion.}
	\label{fig:interplay_exclusion}
\end{figure}

It is interesting to compare our simplified model results with those obtained using the EFT framework. To that end, we show  in Fig.~\ref{fig:interplay_exclusion} the effective coupling $g_{\mathrm{eff}} = g_L = g_Q = C_1^{1/4}$ versus the scale of new physics $\Lambda = (m_S^4\,m_F)^{1/5}$ for the different mass hierarchies shown in the previous figures. Note that we fix the absolute masses $m_F, m_S$ only indirectly via the scale $\Lambda$. We indicate the limit on the scale of new physics when naively assuming $g_{\mathrm{eff}} = g_L = g_Q = 1$ with a red star. Comparing the three different panels, it becomes again obvious that whether $0\nu\beta\beta$ decay or collider searches are more sensitive crucially depends on the relative mass hierarchy. As $0\nu\beta\beta$ decay is cannot resolve the heavy new physics, combining both experimental approaches is crucial.\\ 

\begin{figure}[t]
	\centering
	\includegraphics[width=\textwidth]{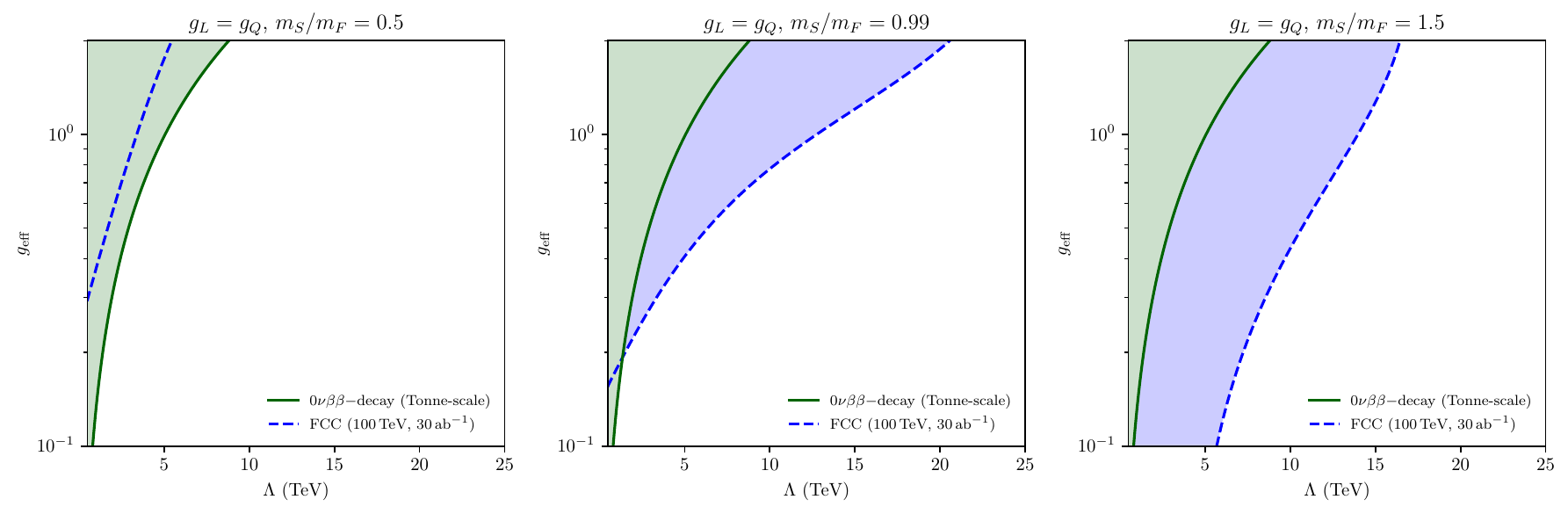}\\
	\caption{
	Complementarity between the next generation of colliders and $\onbb-$decay experiments for $g_L=g_Q$ and $m_S/m_F=\{0.5,0.99,1.5\}$. The definitions of $\Lambda$ and $g_\mathrm{eff}$ are the same as in Fig. \ref{fig:interplay_exclusion}.
	In dark green, we show the $\onbb-$decay exclusion from future tonne-scale experiments, and the reach the FCC-hh LNV searches at 100 TeV with 30 ab$^{-1}$ is shown in blue.}
	\label{fig:interplay_FCC_reach}
\end{figure}

In Fig.~\ref{fig:interplay_FCC_reach}, we finally show a similar set of plots, in which we allow for a larger new physics scale $\Lambda$, demonstrating the future reach of a tonne-scale $0\nu\beta\beta$-decay experiment and the FCC-hh. While $0\nu\beta\beta$-decay experiments will reach a sensitivity of $\Lambda \approx 10~\mathrm{TeV}$, the FCC-hh will reach between  $\Lambda \approx 5~\mathrm{TeV}$ and $\Lambda \approx 15~\mathrm{TeV}$, depending on the mass hierarchy.\\

These results demonstrate that observing a LNV signal at $0\nu\beta\beta$-decay or collider experiments has the potential to exclude the standard thermal leptogenesis scenario. Given their complementary experimental reach, the combination of the high- and low-energy frontier is crucial to probe the mechanism behind the baryon asymmetry generation.\\

\section{Conclusions}
\label{sec:conclusions}
The observation of LNV would have profound implications for our understanding of nature, from the origin of neutrino masses to the matter-antimatter asymmetry of the Universe. In this work, we studied the interplay between three different physical aspects linked by LNV at the TeV scale: $\onbb$-decay, collider phenomenology, and thermal leptogenesis. Previous studies have been performed \cite{Deppisch:2015yqa, Deppisch:2017ecm} from an effective field theory (EFT) standpoint, showing the complementarity between current and future experimental results and their potential to falsify the standard thermal leptogenesis mechanism as an explanation for the origin of matter. Since these first analyses have not yet considered the latest methods, such as the inclusion of background in the collider analysis or the RGE running of the Wilson coefficients across the different energy scales, we extended previous work and studied the impact on the resulting phenomenology. Moreover, we specify a concrete, simplified model that allows us to study aspects that are not possible to capture by a pure EFT approach, such as different mass hierarchies of the new physics involved. By addressing the key questions with a simplified model using state-of-the-art techniques, we also establish a theoretical and computational setup for any other model to perform similar studies like this one.\\

To this end, we considered the SM extended by at least two right-handed neutrinos leading to neutrino masses via the Type I see-saw mechanism and additionally a scalar doublet $S$ and a Majorana singlet $F$, both having masses around the TeV-scale. This model was introduced in Ref. \cite{Peng:2015haa}, where the interplay between collider searches at the LHC and $\onbb$-decay experiments was studied. Besides a detailed analysis of the implications of a possible observation of a $\Delta L = 2$ lepton-number violating signal for the validity of standard thermal leptogenesis, we extended previous works by using the latest hadronic and nuclear matrix elements, improving on the derivation of the $\onbb$-decay half-life, and updating the corresponding predictions.\\

We also revisited the collider study in Ref. \cite{Peng:2015haa}, where a prompt signature of two electrons plus jets in the final state at the LHC was analyzed. There, the different backgrounds (generated using standard MC techniques with misidentification and mistagging probabilities) were differentiated from the signal using a cut-based analysis. Our collider study differs from that of Ref. \cite{Peng:2015haa} in both generation and analysis of events. Firstly, we extended the $\Delta L = 2$ lepton-number violating prompt signature by considering both $e^-e^-$ and $e^+e^+$ in the final state. The background contributions were improved by implementing data-driven methods to emulate the effects of misidentification and mistagging. Finally, the event classification was based on cutting-edge machine learning (ML) algorithms, specifically boosted decision trees and neural networks. Our experience with ML techniques resonates with current literature regarding versatility and implementation time. With these techniques, we identified the already excluded region from the latest LHC runs as well as the future exclusion potential of the high-luminosity LHC or a hypothetical 100 TeV $pp$ collider. For comparison with the reach of $\onbb$ decay, we identified three scenarios with different mass hierarchies between the new particles $S$ and $F$. We demonstrate that collider searches are more sensitive to heavier $S$ due to an enhancement via on-shell production.\\

To study the effect of these new TeV-scale interactions in the context of thermal leptogenesis, we have implemented a set of Boltzmann equations, including the usual expressions involving RHNs \cite{Giudice:2003jh, Buchmuller:2004nz}. In order to account for the washout processes arising from the new interactions of our model, we extended this set-up by the corresponding rates of the relevant LNV terms. We have studied the implications of these new interactions for the weak and strong washout regimes. For both of them, we have found that for couplings larger than $g_L \approx \mathcal{O}(10^{-4})$ or $g_L \approx \mathcal{O}(10^{-6})$, depending on new particle mass hierarchy, an observation of a TeV-scale interaction implies a fast enough washout of any asymmetry previously generated by the out-of-equilibrium decay of RHNs at a high scale, as shown in Fig. \ref{fig:LG_allowed_forbidden}. The extension of the viable thermal leptogenesis region is dependent on the new particle spectrum. This crucial feature was not readily apparent within the pure EFT approach previously employed, emphasizing the advantage of our current methodology in uncovering essential aspects of the thermal history.\\

Based on our analysis, we could demonstrate that the discovery of LNV via the observation of $0\nu\beta\beta$ decay could preclude the viability of standard thermal leptogenesis if TeV-scale LNV interactions dominate the decay process. Our results are generally consistent with previous EFT estimates \cite{Deppisch:2015yqa, Deppisch:2017ecm, Chun:2017spz}. However, in order to confirm that the dominant contribution arises from a dim-9 contribution, additional information is needed. Different ideas for identifying the underlying mechanism exist, such as comparison of results from different isotopes \cite{Bilenky:2004um, Deppisch:2006hb, Gehman:2007qg}, observation of a discrepancy with the sum of neutrino masses determined by cosmology~\cite{DellOro:2015kys}, a deviation in meson decays~\cite{Li:2019fhz, Deppisch:2020oyx}, or signals of LNV TeV-scale new physics from $pp$ collider searches. Besides being able to possibly confirm the underlying new physics, the observation of an LNV signal at the Large Hadron Collider and/or a future 100 TeV $pp$ collider would independently render standard thermal leptogenesis invalid. We could demonstrate that the relative potential of $0\nu\beta\beta$-decay and collider experiments to falsify the standard thermal leptogenesis scenario depends decisively on the new particle spectrum. We would like to stress that the observation of such an experimental signature would not necessarily be in conflict with the scale of light neutrino masses implied by neutrino oscillation experiments as well as cosmological and astrophysical neutrino mass probes. Our analysis also brings to light the opportunities offered by the relative smallness of LNV couplings, essential for a viable leptogenesis scenario. This aspect particularly opens a door for collider signatures of long-lived particles (LLP). New strategies to seize these opportunities are already being developed and will be further explored in our future work.\\

While our analysis was focused on the generation of a lepton asymmetry at a high scale via the decay of right-handed neutrinos, the general implications can be, in principle, transferred to similar mechanisms such as other high-scale leptogenesis scenarios. Specific models, however, might escape the general implications, e.g., scenarios with a dark sector featuring a global $U(1)_X$ symmetry \cite{Frandsen:2018jfi}. Therefore, in order to conclusively falsify models, a dedicated analysis should be performed. Moreover, while we concentrated in our analysis on the electron sector only, there is the caveat that a lepton asymmetry was generated in another (decoupled) flavor sector. In order to address this point, we plan to extend our work by studying flavor effects, which open up interesting links to new collider signatures and low-energy observables.\\

As the discovery of a TeV-scale LNV signal at $\onbb$-decay experiments or current and future colliders will have far-reaching consequences on the validity of standard thermal leptogenesis, such searches are of high relevance in the quest for new physics and, in particular, for the origin of the baryon asymmetry of our Universe.\\

\begin{acknowledgments}
MJRM thanks G. Li for many useful discussions regarding the hadronic and nuclear matrix elements. MJRM, SUQ, and TYS were supported in part under the U.S. Department of Energy contract DE-SC0011095. MJRM was also supported in part under the National Natural Science Foundation of China grant No. 19Z103010239. SUQ thanks B. Shuve, J. de Vries, and S. Krishnamurthy for fruitful discussions. The work of SUQ was supported by ANID-PFCHA/DOCTORADO BECAS CHILE/2018-72190146. JH acknowledges support from the DFG Emmy Noether Grant No. HA 8555/1-1.
\end{acknowledgments}

\appendix
\section{Machine Learning techniques in Collider Analysis}

Differentiating the signal ($S$) from the background ($B$) is a typical classification problem that can be solved using machine learning techniques. Given an ensemble of observables $X$, for each collider event, one can train a model $\mathcal{M}$ to separate signal events from background events with high accuracy. In this paper, we primarily use a recurrent neural network (RNN) to train the classification and a boosted decision tree (BDT) to cross-check the performance of our discriminant.\\

\subsection{Boosted Decision Tree (BDT)}
\begin{figure}
\centering
 \includegraphics[width=0.65\textwidth]{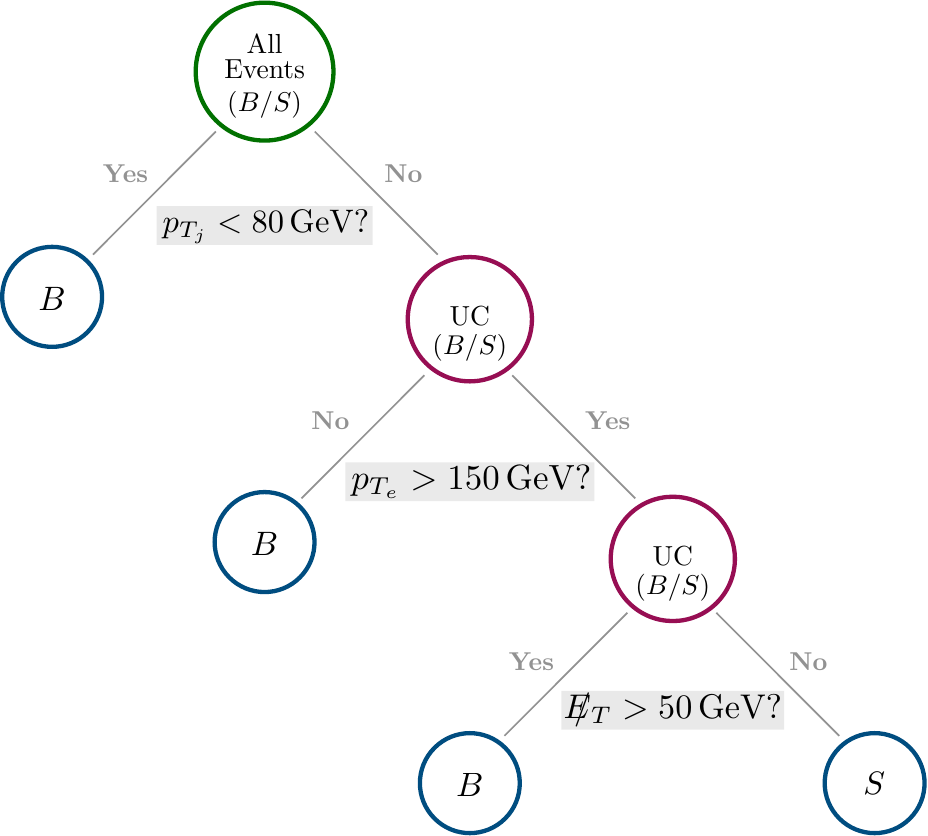} \\
 \caption{A simple 3-depth decision tree example, in which each note represents a cut based on the input parameters ($p_{T_j}$, $p_{T_e}$, and $E\!\!\!/_T$). ``UC'' standards of ``unclassified"  in the picture. Notice this is only an illustration of a decision tree, and it is not reflecting the true classification result of our collider study. }
 \label{fig:DT}
\end{figure}

A decision tree is a set of criteria in a tree-based structure that recursively splits the events into two groups. Following the simplified diagrammatic representation shown in Fig. \ref{fig:DT}, one can start with a set of unclassified events. At each node, the criterion is defined such that ``background-like" events are removed, and this continues until the  signal events are efficiently separated from the background. An ensemble algorithm such as boosting can be applied to this decision tree to further improve the classification, and this forms the BDT.\\

At each node split, the $S$ and $B$ separation can be improved further by using certain criteria such as the Gini index and entropy factor. These criteria are defined such that minimizing them at each node increases the purity of the $S$ and $B$ data sets, hence maximizing the discriminating power. The detailed mathematical definition of the Gini index and entropy could be found in any machine learning textbook.\\

The decision tree method is powerful but can be easily over-fitted, i.e., a tiny change in the input data set may result in large differences and inconsistencies in the classification results. To avoid this, a set of BDTs can be trained in a sequence such that the successive tree is created to minimize the error of the previous tree. Several sets of these can be trained, and the final classification result is determined by the majority vote from all the BDTs. In this study, we use the AdaBoost\footnote{AdaBoost is a method implemented by Toolkit for Multivariate Data Analysis (TMVA), a built-in package in ROOT \cite{Hocker:2007ht}.} (Adaptive Boost) to test a small part of the parameter space of our simplified model.\\

The details of the algorithm are described as follows. The $i-$th tree trained is called $T^i$, $W^i$ is its voting weight, and $w_j^{i}$ is the weight of the $j-$th sample in the $i-$th tree:

\begin{itemize}
	\item
	The first tree ($T^1$) is trained such that all the samples have the same weight. Either the Gini or the entropy criteria can be used for this training. For each sample, the tree predicts if the event is ``signal-like" or ``background-like". The discrete output for each event is as follows:
	$$y_j^i = \left\{\begin{array}{cl} +1 & \textrm{for signal-like events,} \\ -1 & \textrm{for background-like events.}\end{array}\right. $$
	
	\item
	Each tree, $T^i$, has the voting weight $W^i$ which is defined as :
	\begin{equation}
	W^i = \frac{1}{2}\log\left(\frac{1-\it err}{\it err}\right)
	\label{eq:BDTW}
	\end{equation}
	
	with $\it err$ being the ratio of the misclassified samples to the total samples. A sample is misclassified if its prediction $y_j^i$ is different than the truth value $\hat{y}$, where $\hat{y}$ is 1 for signal events and -1 for background events. The voting weight is higher for trees with better classification. A small shift inside the logarithm will be added in practice to prevent infinity when the error is 1 or 0.\\
	
	\item
	Subsequent trees ($T^2$, $T^3$, \ldots, $T^N$) are generated. For each tree, $T^i$, that is  trained, every sample is re-weighted depending on how it was classified. If the classification is correct, the sample weight is given as $w_j^{i} = w_j^{i-1} e^{-W^{i-1}}$. If the classification is incorrect, then the sample weight is given by $w_j^{i} = w_j^{i-1} e^{W^{i-1}}$. After the training and the re-weighting, all weights ($w_j^i$) will be normalized, meaning for each tree, the weights ($w_j^i$) will sum to 1.
	
	\item
	In each of the training, when certain samples are misidentified by the tree, they would be emphasized in the next tree due to the re-weighting. This is because misidentified samples from the previous tree would have a higher weight in the next training, forcing the Gini or entropy criteria to classify them correctly.
	
	\item
	The process of tree training will end when a previously determined total number of trees $N$ has been reached. 
	
	\item
	The final classification result, $y_j$ for the $j-$th sample is 
	\begin{equation}
	y_j = \frac{1}{N}\sum_{i=1}^N{y_j^i\,W^i}\ ,
	\end{equation} 
	where $W^i$ is the voting weight of the $i-$th tree and $N$ is the total number of trees that are trained. 
\end{itemize}

This BDT method is a powerful algorithm for signal-background classification however is very time-consuming. This is because each tree must be generated sequentially, and a completely new series of trees must be trained with new parameter choices for the simplified model. Given that we are interested in several different choices of masses and couplings in the model, we implement a different method described in the next section. This method allows us to build a classification model for the whole parameter space efficiently.

\subsection{Recurrent Neural Network (RNN)}
\label{app_RNN}

A neural network (NN) is a deep learning model comprising of a series of linear and non-linear transformations. The goal is to find an optimal set of parameters that transform a set of initial inputs to approximate the target. The idea is that this predictive model is made of connected units or nodes that mimic the neurons in the brain. The network consists of a series of layers that ``learn" to classify the events as signals or backgrounds through transformations. The type of the layer and the number of layers are defined by the network topology and are optimized to improve the classification. When we provide the network with a set of inputs, $x$, it passes through these layers, undergoing transformations, one after the other. The network then outputs its set of predictions $y$.\\

To improve the NN, one can define a loss function $\mathcal{L}(y,\hat{y})$, where $\hat{y}$ is the truth value and $y$ is the model output. As the NN becomes a better classifier, the predictions $(y)$ will get closer to the truth value $(\hat{y})$ hence reducing the loss. Hence one can optimize the parameters inside the NN by minimizing the loss function.\\ 

\begin{figure}[ht]
	\centering
	\includegraphics[width=0.85\textwidth]{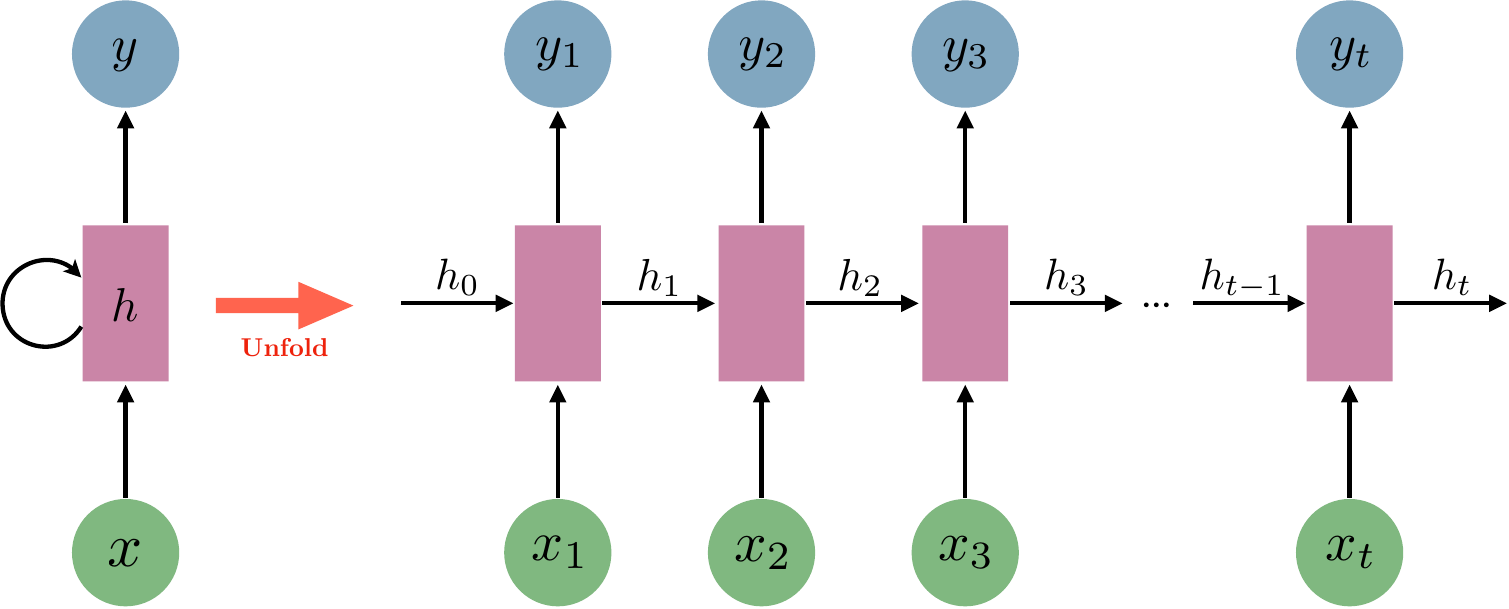}\\
	\caption{Visual representation of a standard RNN in folded version (left) and unfolded version (right). The sequence $\{x_1, x_2, x_3, \ldots,x_t\}$ represents the input, $\{ y_1, y_2, y_3,\ldots,y_t\}$ represents the predicted output, and $\{h_0, h_1, h_2,\ldots, h_t\}$ holds the information from the previous input. The graph illustrates that, at any given time, $t$, the current layer will be updated with respect to a new input.}
	\label{fig:RNNgraph}
\end{figure}

Usually, the inputs of a NN have fixed lengths. However, our simulations contain events with different numbers of particles, and the inputs do not have the same length. We can use a recurrent neural network (RNN) as our deep learning tool as it allows inputs of variable lengths using Gated Recurrent Units\footnote{When the RNNs get very deep, they may tend to suffer from two major weaknesses - divergent or vanishing gradients during the minimization of the loss function. In simple words, the network has difficulties in ``learning" from inputs far away in the sequence, and it makes predictions based mostly on the most recent ones. A typical manner to address this problem is by using a GRU \cite{Guest:2018yhq}.} (GRUs) \cite{DBLP:journals/corr/ChoMGBSB14, DBLP:journals/corr/ChungGCB14}. A standard RNN has the property that the structure of the hidden layers will be updated when new inputs are provided, and it also has the ability to ``remember" parts of the previous input for optimized classification. This is depicted in Fig. \ref{fig:RNNgraph}.\\

\begin{figure}[ht]
	\centering
	\includegraphics[width=0.85\textwidth]{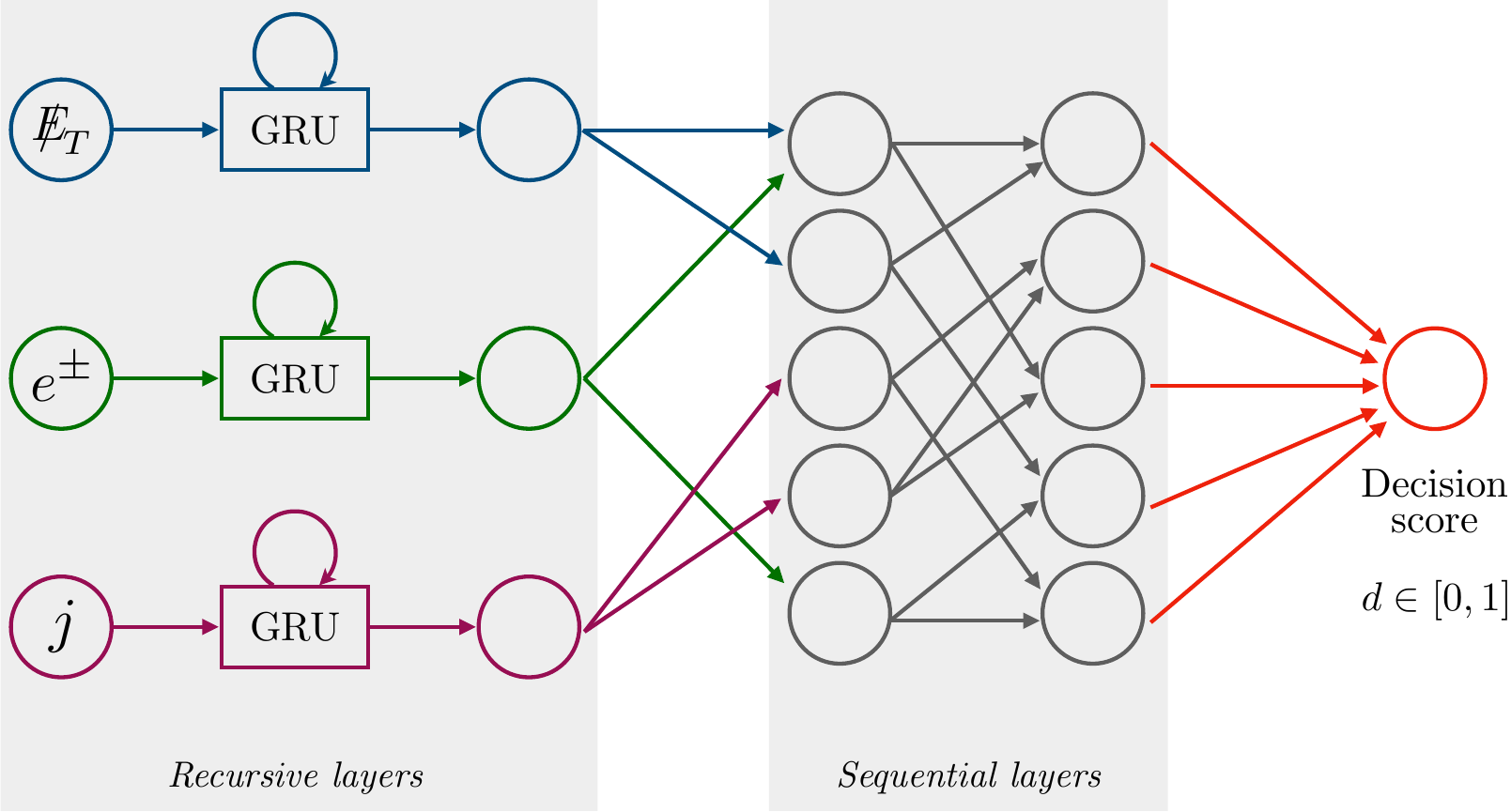}\\
	\caption{Visual representation of the Recurrent Neural Network (RNN) used in our study. The kinematic properties of jets, electrons, and missing energy are analyzed by independent Gated Recurrent Units (GRUs). The outcome is merged into a sequential neural network that produces a decision score $d$ between 0 and 1.}
	\label{fig:RNN}
\end{figure}

As stated before, we use this RNN to separate signal and background events. Kinematic properties of jets, electrons, and missing $E_T$ are used as inputs to three independent, recurrent networks. These are initially trained in parallel and then merged together into a fully connected sequential neural network. The described topology is depicted in Fig. \ref{fig:RNN}. Based on the input variables, the network assigns a decision score ($d$) to every event. This score goes from $d=0$, indicating a perfectly-background-like event, indicating perfectly-background-like event, to $d=1$, indicating a perfectly-signal-like event. We then determine a cutoff $d^\ast$ such that all events satisfying $d>d^\ast$ are classified as signal $S$, and the rest is background, maximizing the signal significance $S/\sqrt{S+B}$.

\clearpage

\bibliographystyle{apsrev4-1}
\bibliography{O2_LNV}

\end{document}